\documentclass[aps,prd,nofootinbib,superscriptaddress,showkeys,showpacs,a4paper]{revtex4-1}
\usepackage{ulem}
\usepackage{color}
\usepackage{graphicx}
\usepackage{hyperref}   
\usepackage{appendix}
\usepackage{epsfig}
\usepackage{bm}
\hypersetup{colorlinks,linkcolor={blue},citecolor={magenta}}
\usepackage{epstopdf}
\usepackage{amsmath}
\usepackage{subfig}
\usepackage{comment}
\usepackage[dvipsnames]{xcolor}
\newcommand{\be}{\begin{equation}}
\newcommand{\ee}{\end{equation}}
\newcommand{\ba}{\begin{eqnarray}}
\newcommand{\ea}{\end{eqnarray}}

\DeclareMathAlphabet\mathbfcal{OMS}{cmsy}{b}{n}
\begin{document}
\title{Viscous effects of a hot QGP medium in time dependent magnetic field and their phenomenological significance}
%\title{Response of a weakly magnetized hot QCD medium to inhomogeneous electric field: Quasiparticle approach}
\author{Debarshi Dey\footnote{Joint first author}}
\email{debarshi@iitb.ac.in}
\affiliation{Indian Institute of Technology Bombay, Mumbai 400076, India}

\author{Gowthama K K\footnote{Joint first author}}
\email{k$_$gowthama@iitb.ac.in}
\affiliation{Indian Institute of Technology Bombay, Mumbai 400076, India}

\author{Sadhana Dash}
\email{sadhana@phy.iitb.ac.in}
\affiliation{Indian Institute of Technology Bombay, Mumbai 400076, India}

\author{Basanta Kumar Nandi}
\email{basanta@iitb.ac.in}
\affiliation{Indian Institute of Technology Bombay, Mumbai 400076, India}

    \begin{abstract}
In this work, we have studied, for the first time, the impact of a realistic picture of a time dependent electric and magnetic field on the shear and bulk viscosities of the medium. Both the electric and magnetic fields are considered to be exponentially decaying with time, and the study is valid in the regime where the magnetic field strength is weak ($eB\ll T^2$). The evaluation has been done in the kinetic theory framework wherein we have solved the relativistic Boltzmann transport equation within the relaxation time approximation collision kernel. We have shown that the constant weak field results as well as the $B=0$ results in the literature can be obtained as special cases of our general results. We have observed that the shear and bulk viscosities increase with time or equivalently, increase with the decrease of the strength of the magnetic field. To connect these observations with experiments, we have calculated the thermalization time, shear viscosity to entropy ratio ($\eta /s$), and bulk viscosity to entropy ratio ($\zeta /s$).
\end{abstract}	
\maketitle

%%%%%%%%%%%%%%%%%%%%%%%%%%%%%%%%%%%%%%%%%%%%%%%%%%%%%%%%%%%%%%%%%%%%%%%%%%%	
	
\section{INTRODUCTION}	
The quark gluon plasma (QGP) produced in the early stages of the heavy ion collision experiments at Relativistic Heavy Ion Collider (RHIC) and Large Hadron Collider (LHC)\cite{ADAMS2005102} depends critically on the initial conditions of the system such as the temperature, chemical potential and the magnetic field. Recent observations at the RHIC and LHC have indicated the presence of a strong magnetic field in the early stages of heavy-ion collision\cite{PhysRevLett.125.022301,PhysRevLett.123.162301}. This happens when the heavy-ion collisions are non-central, \textit{i.e.} when the ions collide with a finite impact parameter. The nuclei and their constituent partons which interact violently in the collisions, give rise to the QGP medium. Those which do not interact are called the spectator partons and it is these spectator quarks moving away from each other at close to the speed of light that create large magnetic fields, the strength of which could be as large as $eB\sim 10^{-1} m_{\pi}^2 (\simeq 10^{17}\, \text{Gauss})$ for super proton synchrotron energies, $eB\sim m_{\pi}^2$ for RHIC energies, and $eB\sim 15\,m_{\pi}^2$ for LHC energies\cite{Skokov:IJMPA24'2009}.  This strong magnetic field is expected to decay with time. Even though there is no definitive description of the evolution of the generated magnetic field, various models predict that the magnetic field may persist for much longer in the QGP due to a finite electrical conductivity of the medium\cite{PhysRevC.88.024911,MCLERRAN2014184,STEWART2021122308}.  This has to do with the magnetic response of the QGP. For a qualitative understanding, one can resort to semi-classical arguments. The magnetic field created by spectator quarks ($\bm{B^{'}}$) tends  to decay exponentially fast. This decreasing magnetic field induces an electric field circulating around the direction of the magnetic field, in accordance with Faraday's law. In the presence of a conducting medium (QGP), this electric field gives rise to an electric current, which in turn produces a magnetic field ($\bm{B^{*}}$) according to Lenz's law. The direction of $\bm{B^{*}}$ is such that it opposes the decay of $\bm{B^{'}}$, thus, increasing the lifetime of the magnetic field in QGP compared to that in vacuum. This time dependent magnetic field may influence the behavior of the hot QCD matter and hence needs to be taken into account to study the various properties of the QCD medium.
  A large corpus of work exists in the literature where several phenomena associated with the QGP have been studied in the presence of constant electromagnetic fields. This includes transport properties such as electrical and hall conductivities\cite{PhysRevD.96.036009,PhysRevD.99.094031,PhysRevD.100.076016,PhysRevD.95.076008,PhysRevLett.120.162301,PhysRevD.96.114026,PhysRevD.102.114015,Rath:PRD21'2019}, shear and bulk viscosities\cite{Rath2022-yq,PhysRevD.31.53,PhysRevD.96.114027,Rath2021-as,PhysRevD.106.094033}, Seebeck and Nernst coefficients\cite{Dey:PRD102'2020,PhysRevD.103.054024,PhysRevD.99.014015,PhysRevD.102.014030,Zhang2021-eg,PhysRevD.104.076021,Khan:PRC110'2024}, etc. In the presence of a strong magnetic field, the QGP medium may experience chiral imbalance, where the system has an unequal number of right handed and left handed particles and antiparticles. In such a system, an additional current in the direction of the magnetic field is developed. This is called the Chiral Magnetic Effect (CME). Similarly, the Chiral Vortical Effect (CVE) is a chiral  anomaly related effect where a current is developed along the direction of the vorticity. Such exclusively quantum phenomena of CME\cite{KHARZEEV2008227,PhysRevD.78.074033} and, CVE\cite{PhysRevLett.106.062301,KHARZEEV20161} have also been studied considering a strong constant background magnetic field.  Recently, some studies have been undertaken wherein the background magnetic field is taken to be time dependent\cite{K:2023dum,K:2022pzc,K:2021sct,Gowthama:2020ghl}. In this work, we calculate the shear and bulk viscous coefficients of a thermal QCD medium with a time-dependent weak background magnetic field ($eB\ll T^2$). In the previous work\cite{K:2021sct} with time dependent magnetic field, the cyclotron was taken to be equal to the inverse of the decay rate of the magnetic field, for mathematical simplicity. This approximation has been relaxed in this work, which, thus, is more mathematically rigorous. \\

 A system can be driven slightly out of equilibrium due to random fluctuations and/or due to external forces like the electromagnetic fields. The system relaxes back to its equilibrium state by transferring energy and momentum within the system. These transfers of energy and momentum due to the redistribution of the constituent particles in the presence of external fields can be studied macroscopically through the shear and bulk viscosity of the system. Shear viscosity is the reaction of the system to momentum transfer along parallel fluid cells. This arises due the inhomogeneity in the fluid velocity of the hydrodynamical system. Bulk viscosity is the resistance to the volumetric change of the system through either compression or expansion. As the QGP medium expands and cools down it experiences both the shear and bulk viscosities. These viscous coefficients have been critical in establishing the QGP system as a strongly interacting medium. With this, the hydrodynamic modeling of QGP has moved away from ideal hydrodynamics to viscous hydrodynamics where shear and bulk viscous coefficients act as input parameters. In hydrodynamic simulations, different observables, such as the elliptic flow coefficient and the hadron transverse momentum spectrum are largely influenced by the shear and bulk viscosities\cite{SONG2008279,Denicol_2010,PhysRevC.85.044909,PhysRevC.90.034907}. The shear and bulk viscosities have also been helpful in our study of the phase transition from the hadronic medium to the QGP medium as the shear viscosity is a minimum and the bulk viscosity is a maximum  near the phase transition point. The relativistic anisotropic viscosities were first introduced in \cite{PhysRevD.81.045015,HUANG20113075} and a kinetic formalism of these transport coefficients was given in \cite{Denicol:PRD98'2018,Denicol:PRD99'2019}. The presence of a magnetic field breaks the rotational symmetry of the system, therefore the viscous stress tensor is described by five shear viscous coefficients and two bulk viscous coefficients\cite{pitaevskii2012physical,Tuchin_2012,PhysRevD.90.066006,Hernandez2017-vo,HATTORI2019551,PhysRevD.101.056020}. In the presence of the magnetic field, the shear and bulk viscosities have been previously determined using various approaches and techniques, viz. the correlator technique using Kubo formula\cite{HATTORI2019551,PhysRevD.87.114003,PhysRevD.103.096015}, the perturbative QCD approach in a weak magnetic field\cite{PhysRevD.97.056024} and the kinetic theory approach \cite{Pushpa:IJMPE33'2024,Kurian2019-eq}. These coefficients have also been calculated in strongly interacting field theories by employing the AdS/CFT correspondence\cite{Son:AnnRev57_2007,Wondrak:NPA1005'2021,Cai:JHEP09'2008}. In this work, we are going to calculate these viscosities by using the  relativistic Boltzmann transport equation within the relaxation time approximation. \\This might perhaps be the first attempt where the physics of time dependency of the external electromagnetic fields has been incorporated in the analysis of momentum transport in the context of QCD medium. Here we would like to clarify that the time dependency in our formalism appears strictly through the electromagnetic fields and we assume that the medium is static in nature. Our estimations are consistent with the studies so far, in the sense that we have correctly reproduced the earlier results in the literature as special cases of our general result, by taking the proper limits of the time-dependent electromagnetic fields.\\

\textit{ Notations and conventions}:  The subscript  $f$ denotes the quark flavor, \textit{i.e.}, $f=u,\,d$. For gluons, we use the subscript $g$, wherever required. %The quantity $q_{f}$ is the electric charge of the particle with flavor $f$ of the $k$th species. 
The particle velocity is defined as ${\pmb v}_f=\frac{{\bf p}}{\epsilon_f}$,  where ${\bf p}$ is the momentum and $\epsilon_f=\sqrt{p^2+m_f^2}$ is the energy (with $m_f$ as the mass of quark with flavor $f$) of the particle. In the rest of the paper, the flavor index $f$ of the velocity is suppressed. The component of a three vector ${\bf A}$ is denoted with the Latin indices $A^i$. The quantities  $E=|{\bf E}|$ and $B=|{\bf B}|$ denote the magnitude of the electric and magnetic fields.
%%%%%%%%%%%%%%%%%%%%%%%%%%%%%%%%%%%%%%%%%%%%%%%%%%%%%%%%%%%%%%%%%%%%%%%%%%%%

\section{VISCOUS COEFFICIENTS WITH TIME VARYING FIELDS}

The energy-momentum tensor of the QGP fluid can be expressed as,
\begin{equation}\label{1}
    T^{\mu\nu}=\int\frac{d^3p}{(2\pi)^3p^0}\,p^{\mu}p^{\nu}f(x,p),
\end{equation}
where the distribution function $f$ is evaluated from relativistic kinetic theory. In general, the $T^{\mu\nu}$ above, can be decomposed into a free part and a dissipative part,
\begin{equation}
  T^{\mu\nu}=T^{\mu\nu}_{(0)}+\Delta T^{\mu\nu} , 
\end{equation}
where
\begin{align}
    T^{\mu\nu}_{(0)}&=\int\frac{d^3p}{(2\pi)^3p^0}p^{\mu}p^{\nu}f_0(x,p)\label{em},\\
    \Delta T^{\mu\nu}&=\int\frac{d^3p}{(2\pi)^3p^0}p^{\mu}p^{\nu}\delta f(x,p)\label{dem},
\end{align}
$f_0$ is the equilibrium distribution function and $\delta f$ is the infinitesimal deviation of the distribution from equilibrium. The next task is to evaluate $\delta f$, for which we take resort to the relativistic Boltzmann transport equation (RBTE).

In the presence of external Lorentz force, RBTE within the relaxation time approximation is written as,
\begin{equation}
	p^{\mu} \frac{\partial f(x,p)}{\partial x^{\mu}}+ \mathcal{F^{\mu}}\frac{\partial f(x,p)}{\partial p^{\mu}}=
	-\frac{p_{\nu}u^{\nu}}{\tau_R}\delta f(x,p),
	\label{rbte_weak}
\end{equation}
where $f(x,p)=f_0(x,p)+\delta f(x,p)$, with $f_0(x,p)$ being the equilibrium distribution function of  quarks, and $\delta f$ is the infinitesimal deviation of the distribution away from equilibrium, \textit{i.e.}, $\delta f\ll f_0$. Here, $\mathcal{F^{\mu}}=(p^0\bm{v}\cdot \bm{F},p^0\bm{F})$ is a 4-vector that can be thought of as the relativistic counterpart of the classical electromagnetic force with $\bm{F}=q(\bm{E}+\bm{v}\times \bm{B})$ being the background classical electromagnetic force field. Using $F^{0i}=-E^{i}$, and $2F_{ij}=\epsilon_{ijk}B^k$, one obtains, 
\begin{equation}
	\mathcal{F^{\mu}}=qF^{\mu\nu}p_{\nu}.
\end{equation}To solve for $\delta f$ in the first approximation, one replaces all $f$'s by $f_0$'s in the LHS of Eq.\eqref{rbte_weak}. To incorporate the effects of a time varying electromagnetic field, however, it is necessary to retain some $\delta f$ terms in the LHS. Writing the Boltzmann equation out into its components, we have,
\begin{equation}\label{BE}
	\frac{p^{\mu}}{p^0}\frac{\partial  f_0}{\partial x^{\mu}}+ \frac{\partial (\delta f)}{\partial t}+q\bm{E}\cdot \frac{\partial f_0}{\partial \bm{p}}+q(\bm{v}\times \bm{B})\frac{\partial (\delta f)}{\partial \bm{p}}=-\frac{\delta f}{\tau_R}.
\end{equation}
The gluons will satisfy a Boltzmann equation with only the first term in the L.H.S. of the above equation. This is because they are electrically neutral and will not couple to electromagnetic fields. Also, since the time dependence comes exclusively from the external fields, the gluonic $\delta f$ will be time-independent.
We denote the first term in Eq.\eqref{BE} by $L$, and use the following \textit{Ansatz} for $\delta f$ \cite{Feng:PRD96'2017},
\begin{equation}\label{ansatz}
	\delta f=-\tau_R q\bm{E}\cdot\frac{\partial f_0}{\partial \bm{p}}-\bm{\Gamma}\cdot \frac{\partial f_0}{\partial \bm{p}},
\end{equation}
which simply implies,
\begin{equation}\label{df}
	f=f_0-\tau_R q\bm{E}\cdot\frac{\partial f_0}{\partial \bm{p}}-\bm{\Gamma}\cdot \frac{\partial f_0}{\partial \bm{p}}.
\end{equation}
Without loss of generality, we can take $\bm{B}\equiv B\hat{z}$. Further, we take a planar profile for the electric field, $\bm{E}\equiv E_x\bm{\hat{x}}+E_y\bm{\hat{y}}$. Then, using Eq.\eqref{df} in Eq.\eqref{BE}, the Boltzmann equation takes the form,
\begin{equation}\label{BE_mod}
	-\bm{\Gamma}\cdot \frac{\partial f_0}{\partial \bm{p}}-qB\tau_R\left(v_x\frac{\partial f}{\partial p_y}-v_y\frac{\partial f}{\partial p_x}\right)=-L-\tau_R \frac{\partial (\delta f)}{\partial t}.
\end{equation}
%%%%%%%%%%%%%%%%%%%%%%%%%%%%%%%%%%%%%%%%%%%%%%%%%%
\begin{figure*}
    \centering
    \centering
    \hspace{-2.5cm}
    \includegraphics[width=0.485\textwidth]{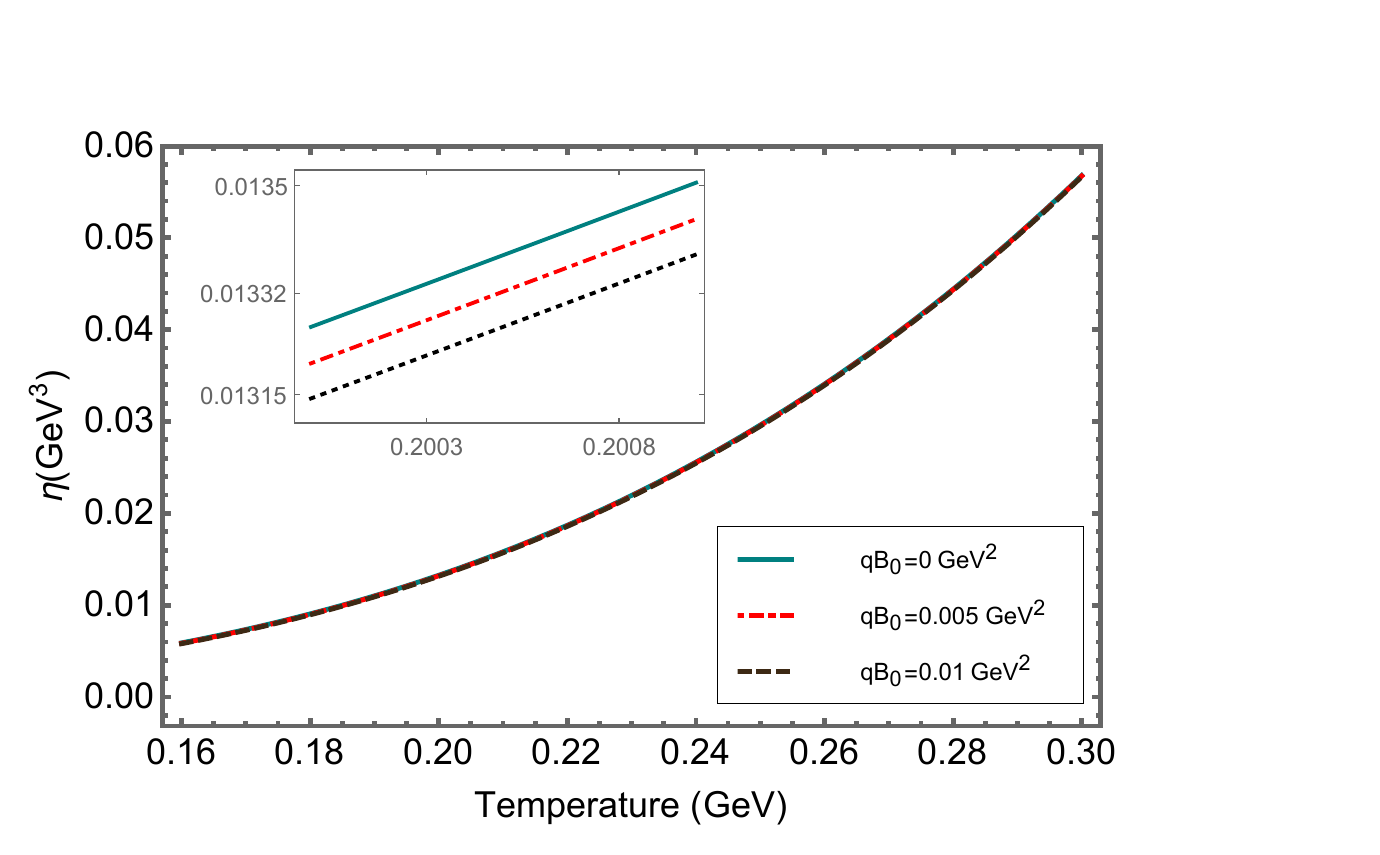}
    \includegraphics[width=0.465\textwidth]{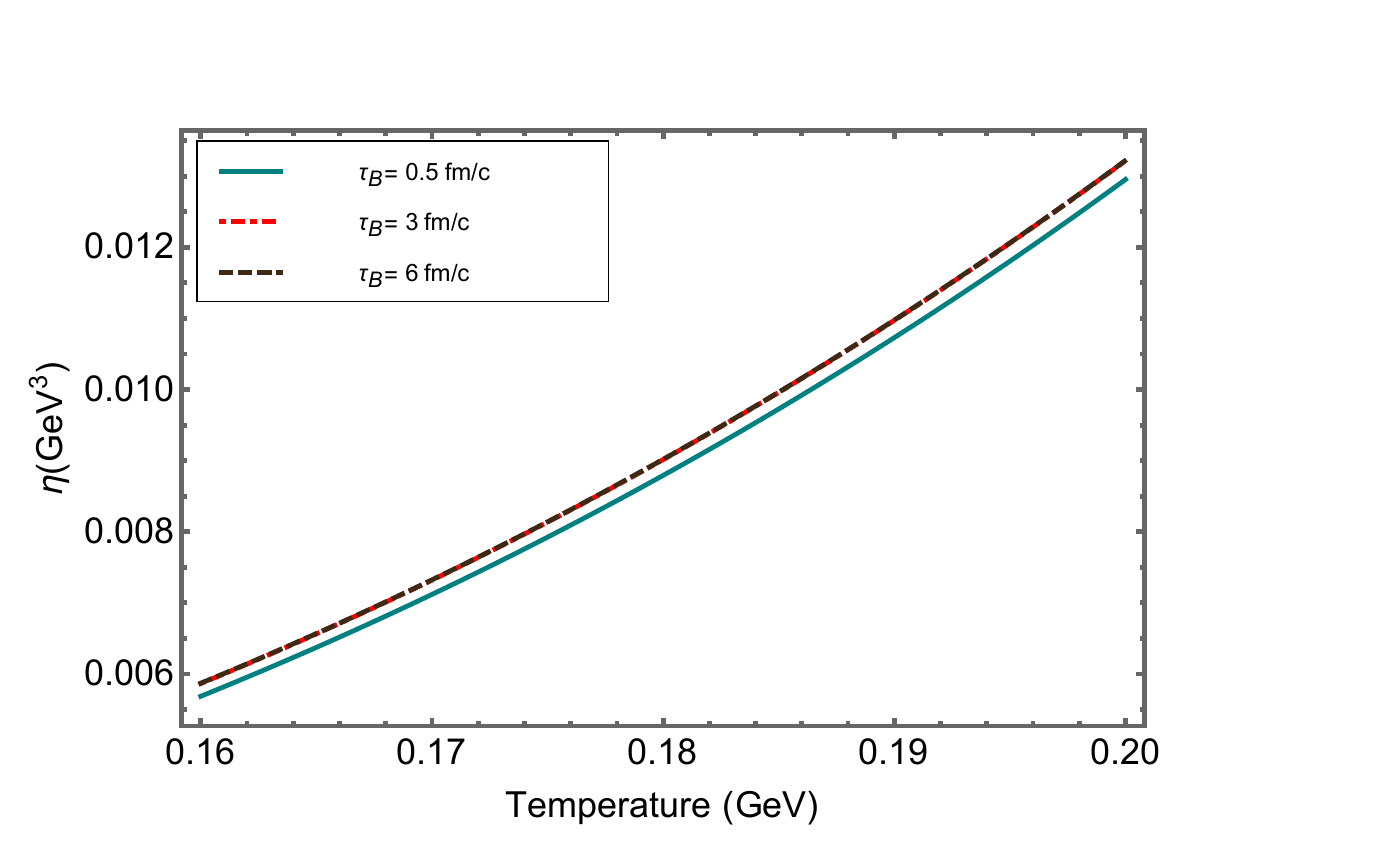}
    \hspace{-2.5cm}
    \captionsetup{justification=raggedright, singlelinecheck=false, format=hang, labelsep=period}
    \caption{{\small Left panel: Temperature behavior of shear viscosity  at various values of the amplitude of magnetic field, $qB_0$, at $t=2$ fm/$c$ and $\mu = 0.01$ GeV. Right panel: Variation of shear viscosity with temperature, and its dependency on the decay rate of the magnetic field, $\tau_B$, with $qB_0=0.01$ GeV$^2$, $t=2$ fm/$c$, $\mu=0.1$ GeV. } }
\label{f1}
\end{figure*}
%%%%%%%%%%%%%%%%%%%%%%%%%%%%%%%%%%%%%%%%%%%%%%%%%%
 We take exponentially decaying electric and magnetic fields given as,\footnote{The electric and magnetic fields defined in Eq.\eqref{ansatz} are external fields which are independent of each other and hence do not satisfy the Maxwell's equations. In general, a time varying magnetic field induces a space dependent electric  field, but as a first approximation, we do not consider the induced electric field in this work.}
\begin{equation}\label{ansatz}
	\bm{B}=B_0e^{-t/\tau_B}\bm{\hat{z}},\qquad \bm{E}=E_0e^{-t/\tau_B}(\bm{\hat{x}}+\bm{\hat{y}}).
\end{equation}
 $\Gamma$ in $\delta f$ [Eq.\eqref{ansatz}] encapsulates the effect of the magnetic field, so, we take $\Gamma$ to be time dependent. Thus, both the terms in Eq.\eqref{ansatz} are time dependent. Now we evaluate the partial derivatives in the parenthesis in Eq.\eqref{BE_mod},
\begin{align}
	v_x\frac{\partial f}{\partial p_y}&=v_x\left\{\frac{\partial f_0}{\partial p_y}-\tau_R qE_x\frac{\partial^2 f_0}{\partial p_y\,\partial p_x}-\tau_R qE_y\frac{\partial^2 f_0}{\partial p_y^2}-\Gamma_x\frac{\partial^2 f_0}{\partial p_y\,\partial p_x}-\Gamma_y\frac{\partial^2 f_0}{\partial p_y^2}-\Gamma_z\frac{\partial^2 f_0}{\partial p_z\,\partial p_y}\right\}, \\
	v_y\frac{\partial f}{\partial p_x}&=v_y\left\{\frac{\partial f_0}{\partial p_x}-\tau_R qE_x\frac{\partial^2 f_0}{\partial p_x^2}-\tau_R qE_y\frac{\partial^2 f_0}{\partial p_x\,\partial p_y}-\Gamma_x\frac{\partial^2 f_0}{\partial p_x^2}-\Gamma_y\frac{\partial^2 f_0}{\partial p_x\,\partial p_y}-\Gamma_z\frac{\partial^2 f_0}{\partial p_x\,\partial p_z}\right\}.
\end{align}
%%%%%%%%%%%%%%%%%%%%%%%%%%%%%%%%%%%%%%%%%%%%%%%%%%
\begin{figure*}
    \centering
    \centering
    \hspace{-2.5cm}
    \includegraphics[width=0.485\textwidth]{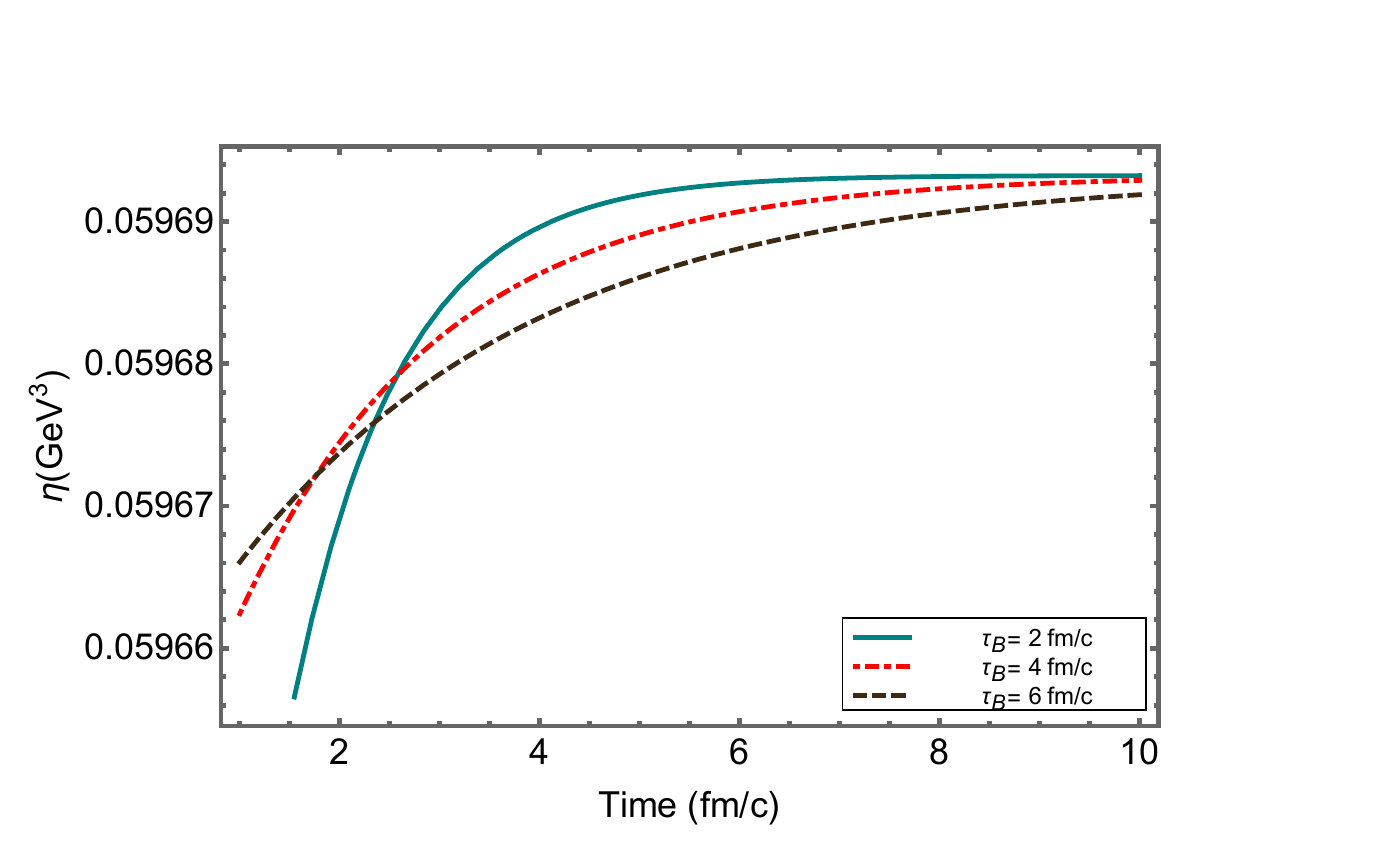}
    \includegraphics[width=0.485\textwidth]{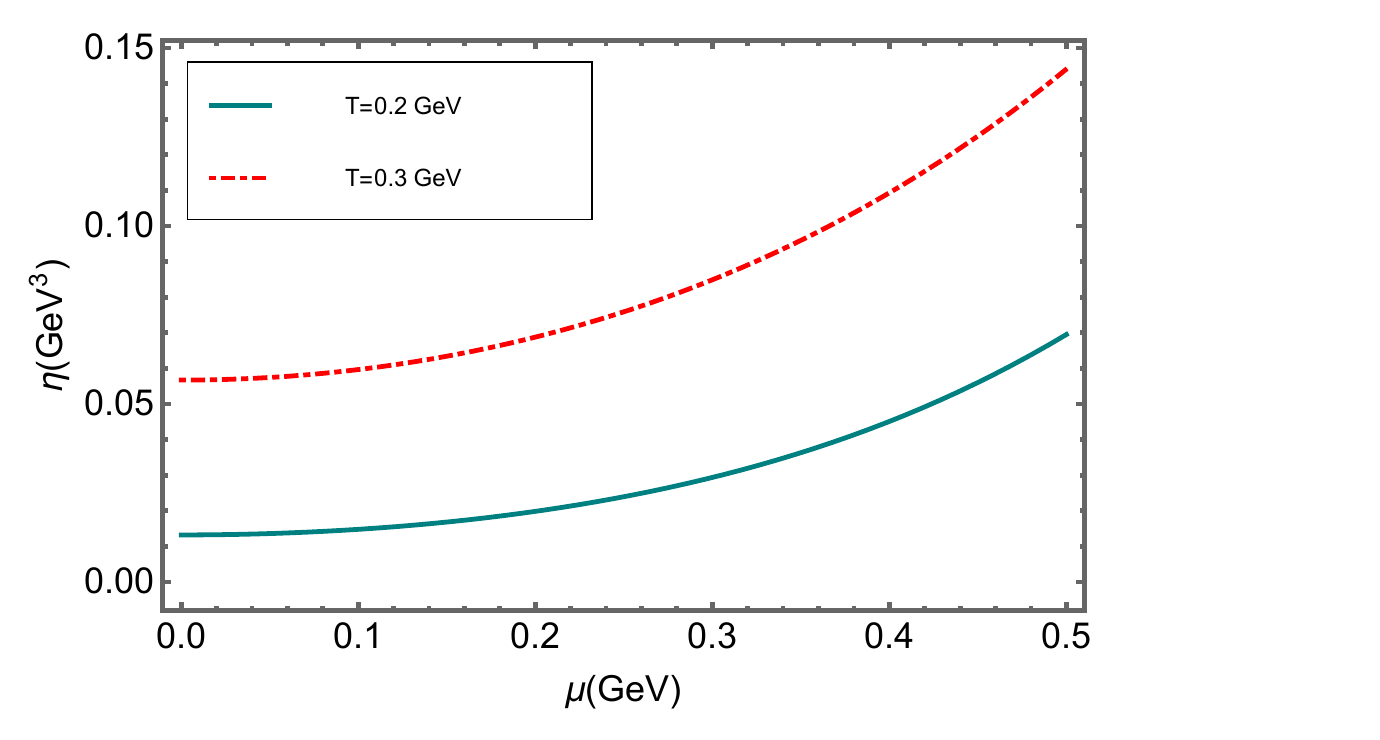}
    \hspace{-2.5cm}
     \captionsetup{justification=raggedright, singlelinecheck=false, format=hang, labelsep=period}
    \caption{{\small Left panel: Shear viscosity as a function of time  for various values of the magnetic field decay rate, $\tau_B$, with $T=0.3$ GeV, $qB_0=0.01$ GeV$^2$, $\mu=0.1$ GeV. Right panel: Shear viscosity as a function of chemical potential $\mu$ at two different values of temperature, with $qB_0=0.01$ GeV$^2$, $\tau_B=4$ fm/$c$, $t=2$ fm/$c$. }}
\label{f2}
\end{figure*}
%%%%%%%%%%%%%%%%%%%%%%%%%%%%%%%%%%%%%%%%%%%%%%%%%%
Considering terms only up to $\mathcal{O}(v)$, we get,
\begin{equation}
	v_x\frac{\partial f}{\partial p_y}-v_y\frac{\partial f}{\partial p_x}  =\frac{\partial f_0}{\partial \epsilon_f}\frac{1}{\epsilon_f}\Big[v_x(-\tau_R q E_y-\Gamma_y)+v_y(-\tau_R q E_x-\Gamma_x)\Big].
\end{equation}	
The Boltzmann equation becomes
\begin{multline}\label{BE2}
	v_x\left[-\Gamma_x+E_y\frac{q^2\tau_R^2B}{\epsilon_f}+\Gamma_y\frac{\tau_R q B}{\epsilon_f}-\tau_R^2q\dot{E_x}-\tau_R \dot{\Gamma_x}\right]\frac{\partial f_0}{\partial \epsilon_f}+\\v_y\left[-\Gamma_y-E_x\frac{q^2\tau_R^2B}{\epsilon_f}-\Gamma_x\frac{\tau_R q B}{\epsilon_f}-\tau_R^2q\dot{E_y}-\tau_R \dot{\Gamma_y}\right]\frac{\partial f_0}{\partial \epsilon_f}+v_z\left[-\Gamma_z-\tau_R\dot{\Gamma_z}\right]+L=0,
\end{multline}	
$L$ is derived in Appendix \ref{AppendixA}. The final expression reads,
\begin{equation}\label{L}
	L=\frac{\partial f_0}{\partial \epsilon_f}\tau_R\Bigg[\left\{\epsilon_f \frac{\partial P}{\partial\epsilon}-\frac{p^2}{3\epsilon_f}\right\}\partial_lu^l+p^l\left(\frac{\partial_lP}{ \epsilon+P}-\frac{\partial_lT}{T}\right)-\frac{Tp^l}{\epsilon_f}\partial_l\left(\frac{\mu}{T}\right)-\frac{p^kp^l}{2\epsilon_f}W_{kl}\Bigg].
\end{equation} 	
This can be written as,
\begin{multline}
L=\frac{\partial f_0}{\partial \epsilon_f}\tau_R\Bigg[\frac{p_0p_xv_x}{p^2}\left\{\epsilon_f \frac{\partial P}{\partial\epsilon}-\frac{p^2}{3\epsilon_f}\right\}\partial_lu^l-\frac{p^kv_xW_{kx}}{2}+\frac{p_0p_yv_y}{p^2}\left\{\epsilon_f \frac{\partial P}{\partial\epsilon}-\frac{p^2}{3\epsilon_f}\right\}\partial_lu^l-\frac{p^kv_xW_{ky}}{2}\\+p^lp^l\left(\frac{\partial_lP}{ \epsilon+P}-\frac{\partial_lT}{T}\right)-\frac{Tp^l}{\epsilon_f}\partial_l\left(\frac{\mu}{T}\right)\Bigg].
\end{multline}	

Here, $l,k=1,2,3$. Using this in Eq.\eqref{BE2}, and equating the coefficients of $v_x$, $v_y$ and $v_z$, we arrive at the following equations:\begin{align}
	\dot{\Gamma_x}+\frac{\Gamma_x}{\tau_R}-\Gamma_y\omega_c+\tau_R\dot{E_x}-q\tau_R\omega_cE_y-L_x&=0,\label{cde1}\\[0.3em]
	\dot{\Gamma_y}+\frac{\Gamma_y}{\tau_R}-\Gamma_x\omega_c+\tau_R\dot{E_y}+q\tau_R\omega_cE_x-L_y&=0,\label{cde2}
\end{align}	
where
\begin{align}
	L_x&=\Bigg[\frac{p_0p_x}{p^2}\left\{\epsilon_f \frac{\partial P}{\partial\epsilon}-\frac{p^2}{3\epsilon_f}\right\}\partial_lu^l-\frac{p^kW_{kx}}{2}\Bigg],\\[0.2em]
	L_y&=\Bigg[\frac{p_0p_y}{p^2}\left\{\epsilon_f \frac{\partial P}{\partial\epsilon}-\frac{p^2}{3\epsilon_f}\right\}\partial_lu^l-\frac{p^kW_{ky}}{2}\Bigg].
\end{align}	
Also, $\omega_c=qB/\epsilon_f$, with $B=B_0\,e^{-t/\tau_B}$.	Eq.\eqref{cde1}, \eqref{cde2} are coupled linear differential equations, which need to be solved for $\Gamma_x$, $\Gamma_y$.
%%%%%%%%%%%%%%%%%%%%%%%%%%%%%%%%%%%%%%%%%%%%%%%%%%
\begin{figure*}
    \centering
    \centering
    \hspace{-2.5cm}
    \includegraphics[width=0.485\textwidth]{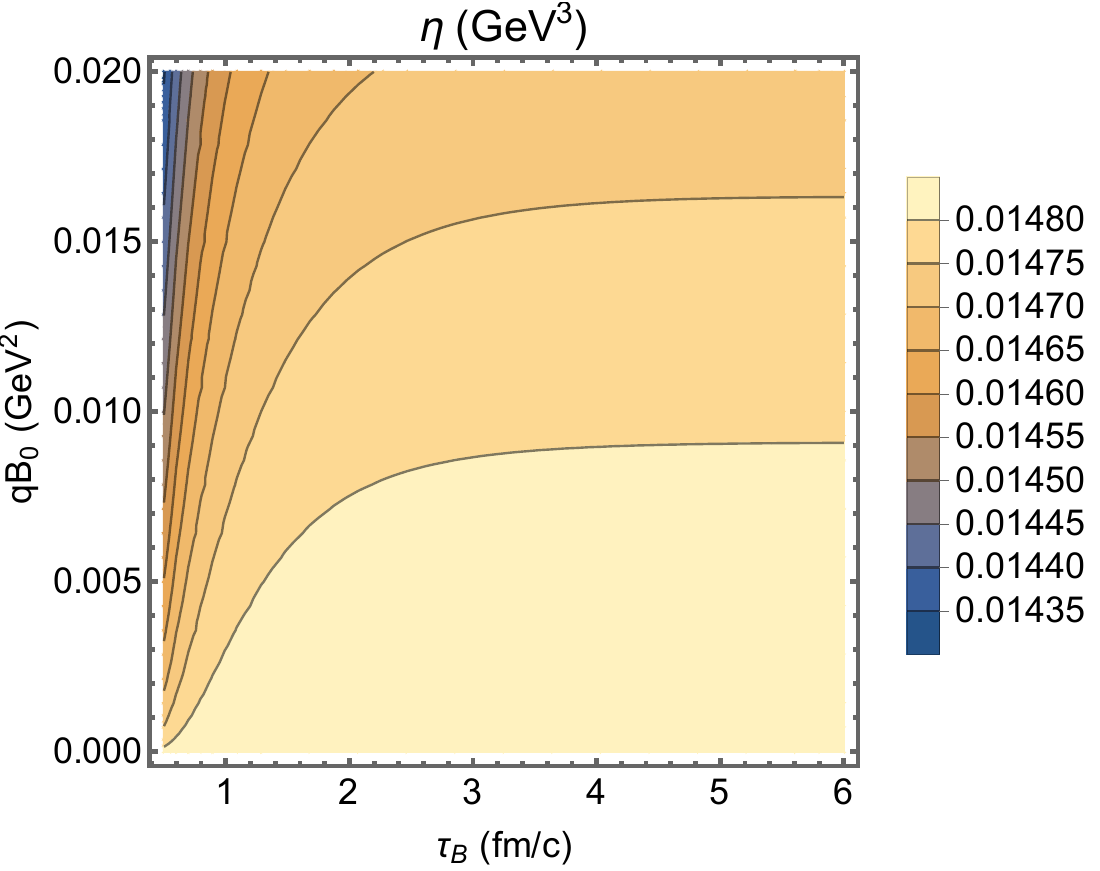}
    \hspace{.5cm}
    \includegraphics[width=0.485\textwidth]{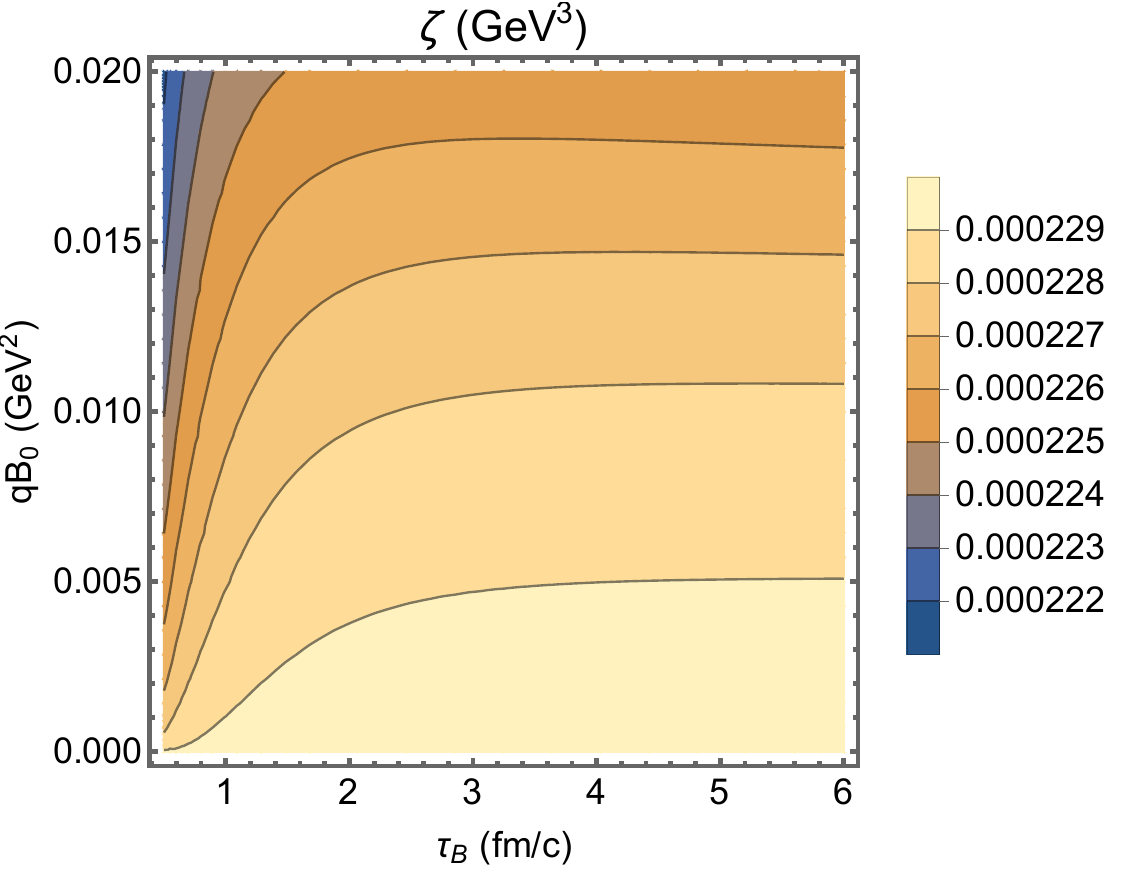}
    \hspace{-2.5cm}
    \captionsetup{justification=raggedright, singlelinecheck=false, format=hang, labelsep=period}
    \caption{{\small Left panel: Contour plot of shear viscosity as a function of the magnetic field amplitude, $qB_0$ and decay rate, $\tau_B$ at $t=2$ fm/$c$, $T=0.2$ GeV and $\mu =0.1$ GeV. Right panel: Contour plot of bulk viscosity as a function of the magnetic field amplitude, $qB_0$ and decay rate, $\tau_B$, at $t=2$ fm/$c$, $T=0.2$ GeV and $\mu =0.1$ GeV.} }
\label{f3}
\end{figure*}
%%%%%%%%%%%%%%%%%%%%%%%%%%%%%%%%%%%%%%%%%%%%%%%%%%

We use the matrix method of solving these equations. Writing the coupled differential equations in matrix form, we have,
\begin{equation}\label{de}
	\frac{d\bm{\Gamma}}{dt}=A\bm{\Gamma}+\bm{S},
\end{equation}
	where 
	\begin{equation}
		\bm{\Gamma}\equiv\begin{pmatrix}
			\Gamma_x\\
			\Gamma_y
		\end{pmatrix},\quad A\equiv \begin{pmatrix}
		-1/\tau_R & \omega_c\\
		-\omega_c & -1/\tau_R
	\end{pmatrix},\quad \bm{S}=\begin{pmatrix}
	-\tau_R q\dot{E_x}+q\tau_R\omega_cE_y+L_x\\
	-\tau_R q\dot{E_y}-q\tau_R\omega_cE_x+L_y.
\end{pmatrix}
	\end{equation}
The corresponding homogeneous equation is,
\begin{equation}\label{deh}
\frac{d\bm{\Gamma}}{dt}=A\bm{\Gamma}.
\end{equation}	
The solution to this equation, called the complementary function, is given by,
\begin{equation}\label{cf}
\bm{\Gamma}\equiv\begin{pmatrix}
	\Gamma_x\\
	\Gamma_y
\end{pmatrix}
=C_1e^{\chi_1}\bm{v_1}+C_2e^{\chi_2}\bm{v_2}.
\end{equation}	

Here, $C_1$ and $C_2$ are constants to be determined from the initial conditions of the problem, with the $\chi$'s given by,
\begin{equation}
	\chi_1=\int dt\,\lambda_1,\quad \chi_2=\int dt\,\lambda_2,
\end{equation}	
where $\lambda_1$, $\lambda_2$ are the eigenvalues of $A$, given by,
\begin{equation}
	\lambda_1=-1/\tau_R-i\omega_c,\quad \lambda_2=-1/\tau_R+i\omega_c.
\end{equation}
$\bm{v_1}$, $\bm{v_2}$ are the eigenvectors of $A$, given by,
\begin{equation}
	\bm{v_1}=\begin{pmatrix}
		1\\-i
	\end{pmatrix},\quad \bm{v_2}=\begin{pmatrix}
	1\\i
\end{pmatrix}.
\end{equation}	
%%%%%%%%%%%%%%%%%%%%%%%%%%%%%%%%%%%%%%%%%%%%%%%%%%
\begin{figure*}
    \centering
    \centering
    \hspace{-2.5cm}
    \includegraphics[width=0.485\textwidth]{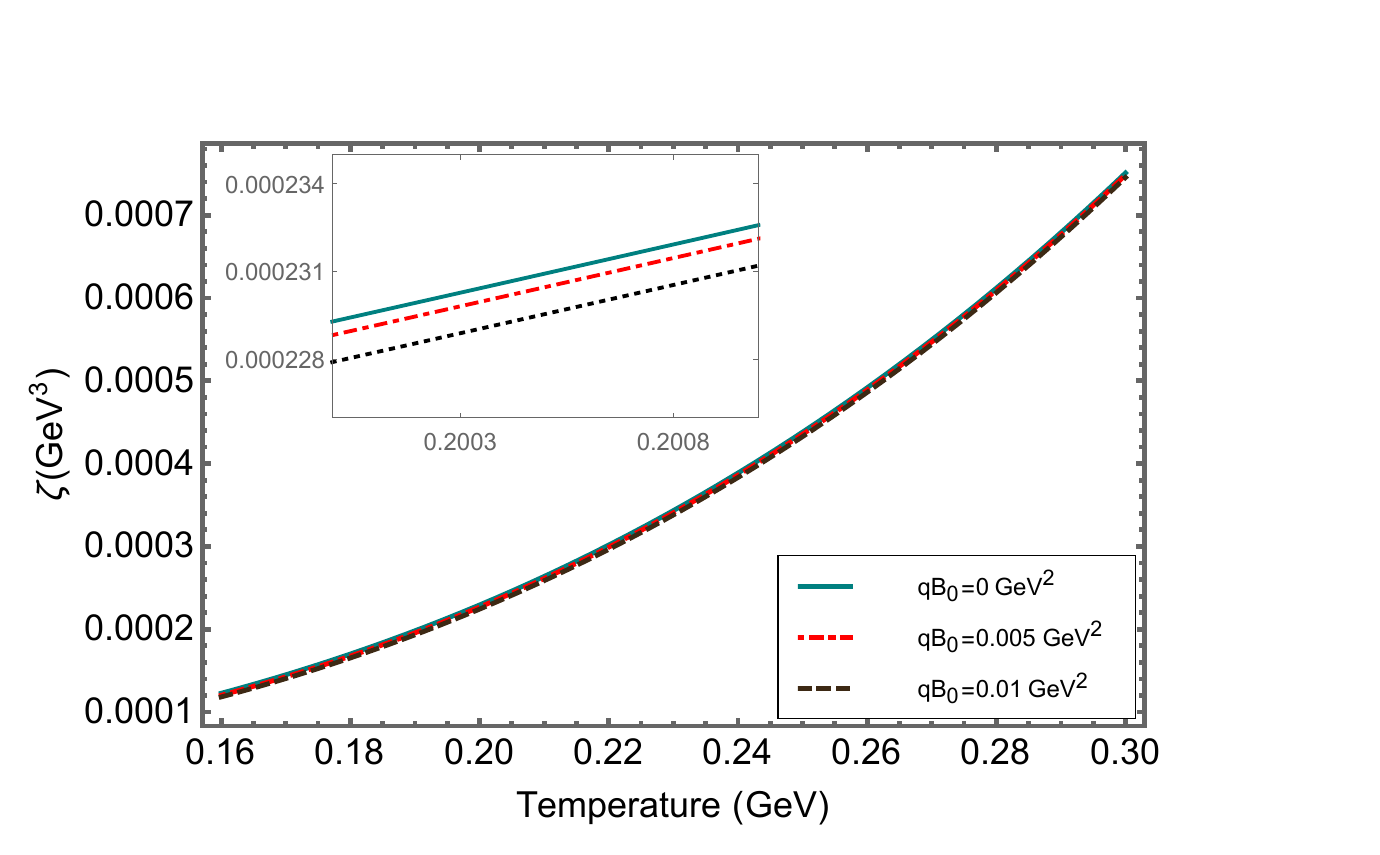}
    \includegraphics[width=0.485\textwidth]{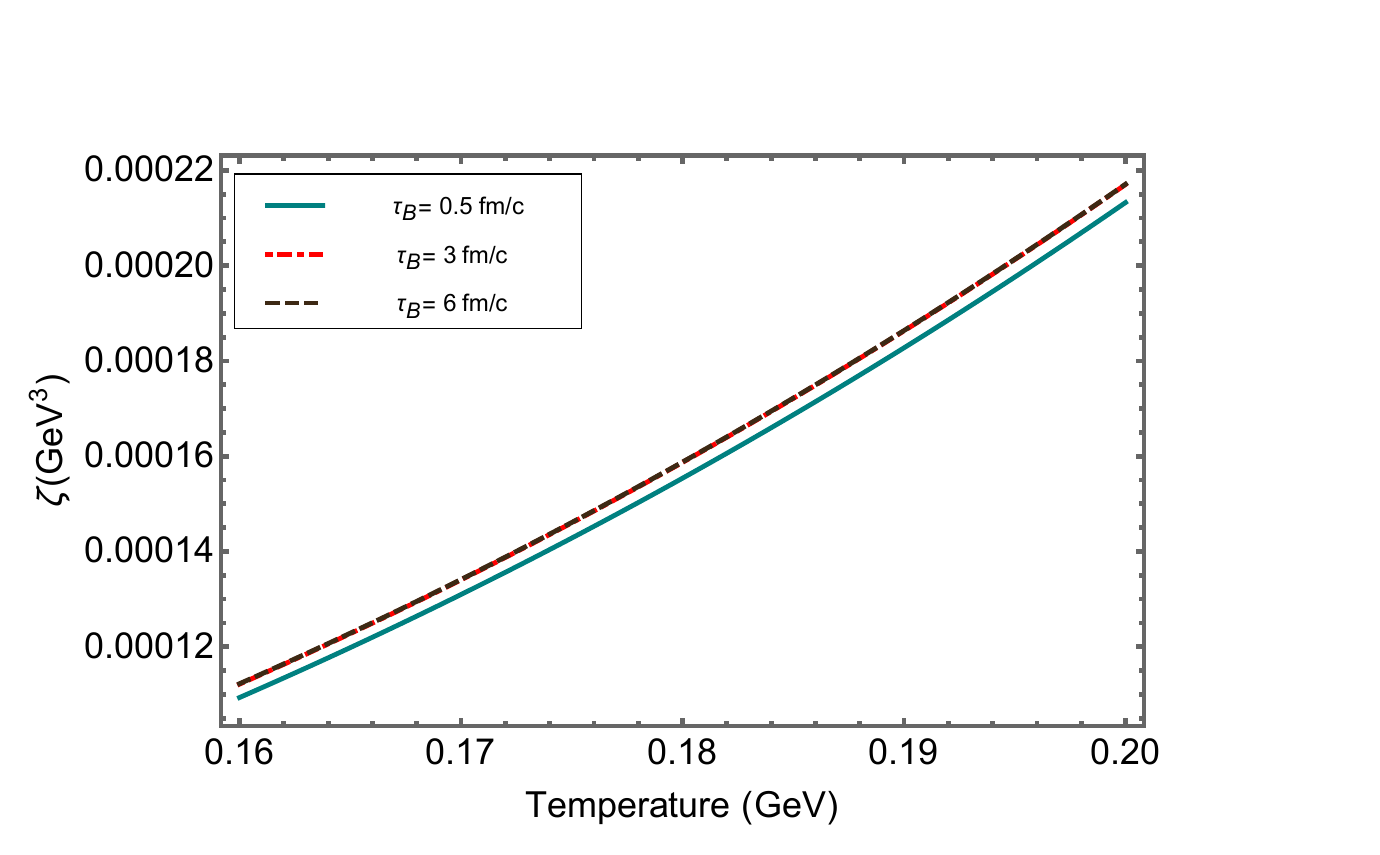}
    \hspace{-2.5cm}
     \captionsetup{justification=raggedright, singlelinecheck=false, format=hang, labelsep=period}
    \caption{{\small Left panel: Temperature behavior of bulk viscosity at various values of the amplitude of the magnetic field, $qB_0$, with $\tau_B=1$ fm/$c$, $t=2$ fm/$c$, $\mu=0.1$ GeV.  Right panel: Variation of bulk viscosity with time for various values of the decay rate of the magnetic field, $\tau_B$, with $qB_0=0.01$ GeV$^2$, $t=2$ fm/$c$, and $\mu = 0.1$ GeV.}}
\label{f4}
\end{figure*}
%%%%%%%%%%%%%%%%%%%%%%%%%%%%%%%%%%%%%%%%%%%%%%%%%%
The particular solution of Eq.\eqref{de} is given by,
\begin{equation}\label{gammap}
	\bm{\Gamma_p}=x(t)\int x(t)^{-1}\bm{S},
\end{equation}	
where $x(t)$ is a square matrix made up of the solutions of the homogeneous equation Eq.\eqref{deh}:
\begin{equation}
	x(t)=\begin{pmatrix}
		e^{\chi_1} & e^{\chi_2}\\
		-ie^{\chi_1} & ie^{\chi_2}
	\end{pmatrix}.
\end{equation} 
$\bm{\Gamma_p}$	is evaluated explicitly in Appendix \ref{AppendixB}. It is a column matrix with two elements, which, we refer to by $\Gamma_{xp}$, $\Gamma_{yp}$.The complete solution to Eq.\eqref{de} is then given by,
\begin{align}
	\Gamma_x&=C_1\exp\left[-t/\tau_R+ibe^{-t/\tau_B}\right]+C_2\exp\left[-t/\tau_R-ibe^{-t/\tau_B}\right]+\Gamma_{xp},\label{gx}\\[0.2em]
	\Gamma_y&=-iC_1\exp\left[-t/\tau_R+ibe^{-t/\tau_B}\right]+iC_2\exp\left[-t/\tau_R-ibe^{-t/\tau_B}\right]+\Gamma_{yp},\label{gy}
\end{align} 
	where $b=\tau_B \frac{ qB_0}{\epsilon_f}$. In the rest of our calculations we are ignoring $C_1$ and $C_2$, which are constants with time but can depend on other quantities such as temperature, $qB_0$ etc. The argument is as follows: In the limiting case of constant field ($t \rightarrow 0$) the particular solutions, $\Gamma_{xp}$ and $\Gamma_{yp}$ exactly reproduce the constant field results of $\eta$ and $\zeta$ in literature, as shown in \eqref{shearconst} and \eqref{bulkconst}. Since switching on the time should not change the contribution from time independent quantities $C_1$ and $C_2$, we can conclude that they will not contribute to the shear and bulk viscosities, and hence,  can be neglected. The elements $\Gamma_{xp}$, $\Gamma_{yp}$ are given by,
 \begin{align}\label{xp1}
   \Gamma_{xp}=&  \left[\int dt\left\{-\frac{\tau_R q}{2}\left(\dot{E_x}+i\dot{E_y}\right)+\frac{q^2\tau_R}{2\epsilon_f}B\left(E_y-iE_x\right)+\frac{1}{2}(L_x+iL_y)\right\}e^{-\chi_1}\right]e^{\chi_1}\,+\nonumber\\
     &\left[\int dt\left\{-\frac{\tau_R q}{2}\left(\dot{E_x}-i\dot{E_y}\right)+\frac{q^2\tau_R}{2\epsilon_f}B\left(E_y+iE_x\right)+\frac{1}{2}(L_x-iL_y)\right\}e^{-\chi_2}\right]e^{\chi_2},
 \end{align}

 \begin{align}\label{yp}
   \Gamma_{yp}=&  \left[\int dt\left\{-\frac{\tau_R q}{2}\left(\dot{E_y}-i\dot{E_x}\right)+\frac{q^2\tau_R}{2\epsilon_f}B\left(-E_x-iE_y\right)+\frac{1}{2}(L_y-iL_x)\right\}e^{-\chi_1}\right]e^{\chi_1}\,+\nonumber\\
     &\left[\int dt\left\{-\frac{\tau_R q}{2}\left(\dot{E_y}+i\dot{E_x}\right)+\frac{q^2\tau_R}{2\epsilon_f}B\left(-E_x+iE_y\right)+\frac{1}{2}(L_y+iL_x)\right\}e^{-\chi_2}\right]e^{\chi_2}.
 \end{align}
	Each element thus has 12 integrals that are to be added to yield the final result. In our analysis we have considered a time varying electric field, but only the terms that come with $L_x$ and $L_y$ contribute to the shear and bulk viscosities. This is because, only $L_x$, $L_y$ contain the velocity gradient structure $W_{ij}$. Hence, all the integrals in Eqs. [\ref{xp1}-\ref{yp}] that do not contain $L_x$ or $L_y$, which are the terms with electric fields, do not contribute to shear and bulk viscosities. The first integral of $\Gamma_{xp}$, is shown below, as an example,
 \begin{equation}
     \int dt\dot{E_x}\,e^{-\chi_1}=-\frac{E_0}{\tau_E}\int dt\, e^{t(-1/\tau_E+1/\tau_R)}\exp\left({-\frac{i\tau_BqB_0}{\epsilon_f}e^{-t/\tau_B}}\right).
 \end{equation}
The integral thus consists of an exponential of an exponential. We make the substitution $u=e^{-t/\tau_B}$ so that the integral becomes,
\begin{equation}
    \frac{E_0\tau_B}{\tau_E}\int du\,\frac{1}{u}\exp\Big(-\tau_B\ln{u}\left[1/\tau_R-1/\tau_E\right]\Big)\exp\left({-\frac{i\tau_BqB_0}{\epsilon_f}\,u}\right).
\end{equation}
	Thus, the first element of $\Gamma_{xp}$ in Eq.\eqref{xp1} is,
 \begin{equation}
    I_1=  \frac{\tau_R q E_0\tau_B}{2\tau_E}\int du\,u^{a-1}\exp({-ibu}),
 \end{equation}
 where
 \begin{equation}
     a\equiv \tau_B\left(\frac{1}{\tau_E}-\frac{1}{\tau_R}\right),\qquad b\equiv \frac{\tau_BqB_0}{\epsilon_f}.
 \end{equation}
The remaining integrals are evaluated in Appendix \ref{AppendixB}.

Evaluating $\Gamma_x$ , $\Gamma_y$ is akin to evaluating $\delta f$ via Eq.\eqref{ansatz}, with $\frac{\partial f_0}{\partial \bm{p}}=\beta \bm{v}f_0(f_0- 1)$.
 %%%%%%%%%%%%%%%%%%%%%%%%%%%%%%%%%%%%%%%%%%%%%%%%%%
\begin{figure*}
    \centering
    \centering
    \hspace{-2.5cm}
    \includegraphics[width=0.487\textwidth]{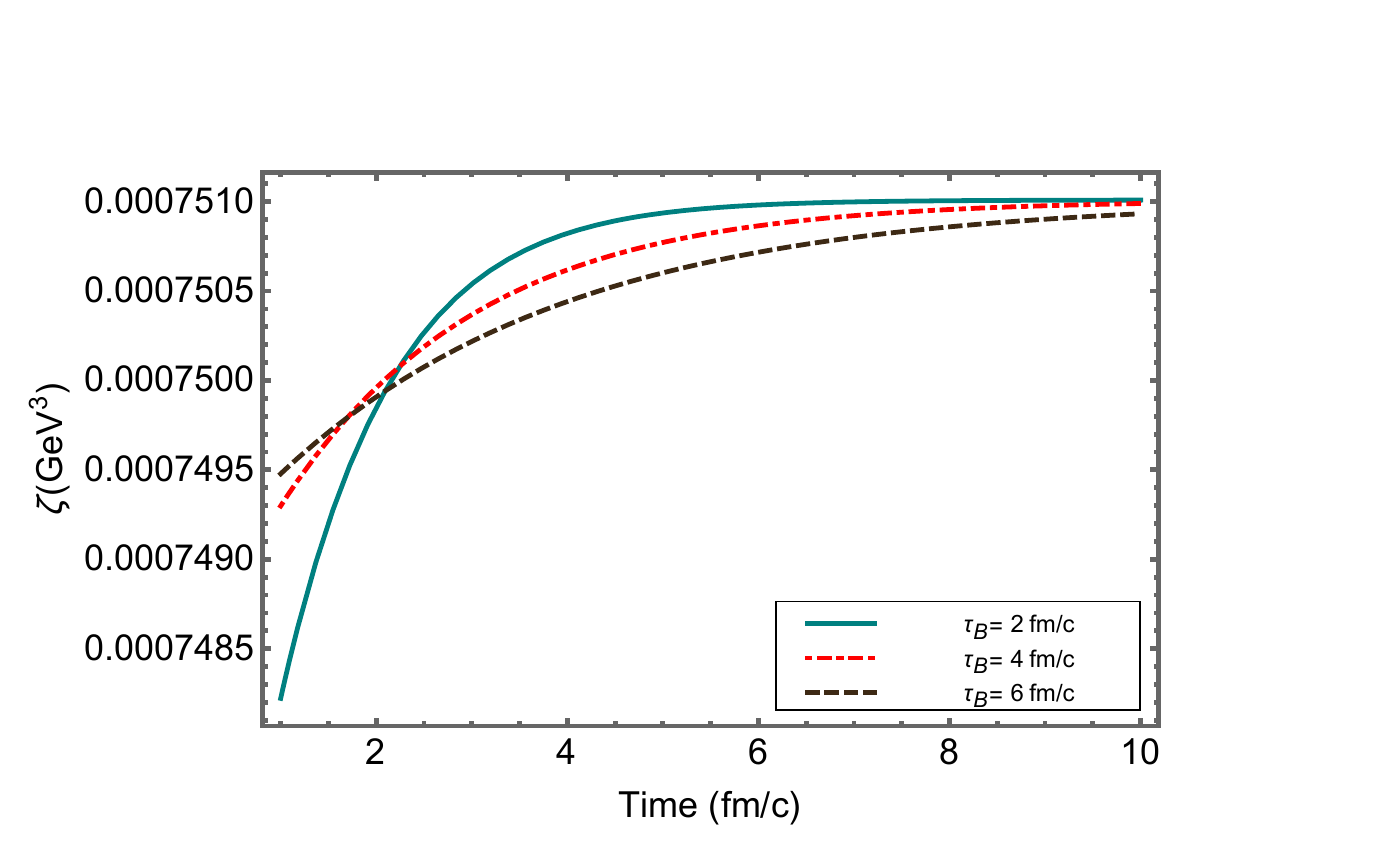}
    \includegraphics[width=0.485\textwidth]{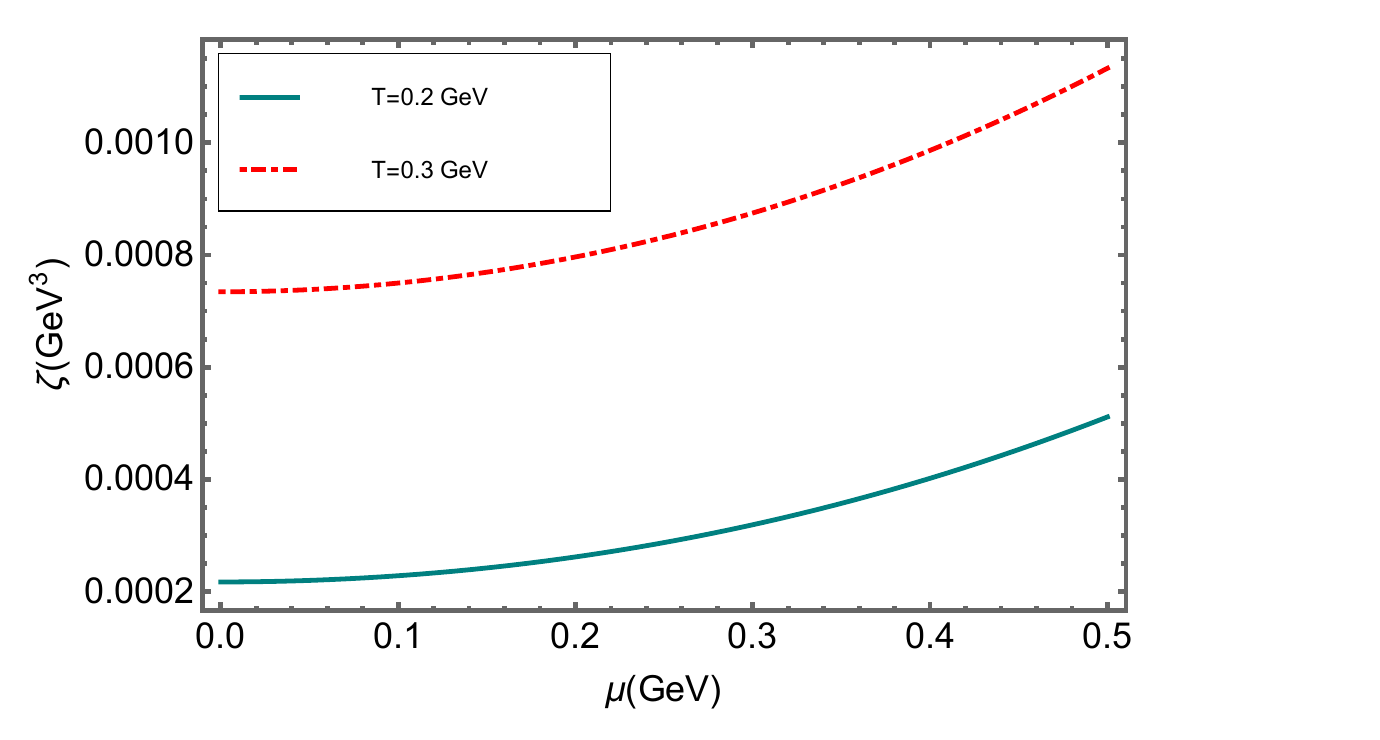}
    \hspace{-2.5cm}
    \captionsetup{justification=raggedright, singlelinecheck=false, format=hang, labelsep=period}
    \caption{{\small Left panel: Bulk viscosity as a function of time for various values of the magnetic field decay rate, $\tau_B$, for $T=0.3$ GeV, $qB_0=0.01$ GeV$^2$, $\mu=0.1$ GeV.  Right panel: Bulk viscosity as a function of chemical potential, $\mu$, for two different values of temperature with $qB_0=0.01$ GeV$^2$, $\tau_B=4$ fm/$c$, $t=2$ fm/$c$.} }
\label{f5}
\end{figure*}
 %%%%%%%%%%%%%%%%%%%%%%%%%%%%%%%%%%%%%%%%%%%%%%%%%%
The spatial components of the dissipative part of the energy momentum tensor can be related to the shear and bulk viscous coefficients in the following manner\cite{pitaevskii2012physical}
\begin{equation}\label{deltaT1}
    \Delta T^{ij}=-\eta W^{ij}-\zeta \delta^{ij}\partial_l u^l,
\end{equation}
where $i,j,l=\{1,2,3\}$, and
\begin{equation}
    W^{ij}=\partial^iu^j+\partial^ju^i-\frac{2}{3}\delta^{ij}\partial_lu^l
\end{equation}   
To obtain $\Delta T^{ij}$ for our problem, we substitute $\delta f$ into Eq.\eqref{dem} and set $\mu,\nu=i,j$, respectively. Further,  we need to consider only those terms in $\delta f$ which possess the relevant tensor structures as in Eq.\eqref{deltaT1}. The details have been mentioned in Appendix  \ref{AppendixC}. The final expression reads,
\begin{multline}\label{deltaT2}
 \Delta T^{ij}=-\frac{1}{2}\Bigg[\sum_fg_f\int\frac{d^3p}{(2\pi)^3}\frac{p^{i}p^{j}}{\epsilon_f}   \bigg\{ \left(\frac{\partial f^0}{\partial \epsilon_f}\right)\left(A_1 e^{\chi_1}+A_2 e^{\chi_2}\right)+\left(\frac{\partial \overline{f^0}}{\partial \epsilon_f}\right)\left(\overline{A_1}e^{\overline{\chi_1}}+\overline{A_2}e^{\overline{\chi_2}}\right)\bigg\}\times \bigg\{\left(\epsilon_f\frac{\partial P}{\partial \epsilon}-\frac{p^2}{3\epsilon_f}\right)\partial_lu^l-\\\frac{p^kp^lW_{kl}}{2\epsilon_f} \bigg\}\Bigg] +g_g\int\frac{d^3p}{(2\pi)^3}\frac{p^{i}p^{j}}{\epsilon_g}\tau_g\beta f_g^0(1+f_g^0)\times \left\{\left(\epsilon_g\frac{\partial P}{\partial \epsilon}-\frac{p^2}{3\epsilon_g}\right)\partial_lu^l-\frac{p^kp^lW_{kl}}{2} +p^l\left(\frac{\partial_lP}{\epsilon+P}-\frac{\partial_lT}{T}\right)\right\},
\end{multline}

where
\begin{equation}
    A_1=\int dt\,e^{-\chi_1},\quad A_2=\int dt\,e^{-\chi_2},\quad \overline{A_1}=\int dt\,e^{-\overline{\chi_1}},\quad \overline{A_2}=\int dt\,e^{-\overline{\chi_2}}.
\end{equation}
The antiquark contribution is taken explicitly in Eq.\eqref{deltaT2} and, therefore, the quark degeneracy factor is equal to six ($g_f=6$). The gluon degeneracy factor, $g_g=16$.
\subsection{Shear Viscosity}
To obtain the shear viscosity, we compare Eq.\eqref{deltaT2} with Eq.\eqref{deltaT1}, and read off the coefficient of $-W^{ij}$. The expression reads,
%\begin{equation}\label{eta}
%    \eta =-\left[\sum_f {g_f} \left(\frac{2}{15}\right)\int \frac{d^3 p}{(2 \pi)^3} \frac{p^4}{2\epsilon_f^2} \left(\frac{\partial f^0}{\partial \epsilon_f}\right)(A_1 e^{\chi_1}+A_2 e^{\chi_2})+g_g\left(\frac{2}{15}\right)\int \frac{d^3 p}{(2 \pi)^3} \frac{p^4 \tau_g}{2\epsilon_g^2} \left(\frac{\partial f^g}{\partial \epsilon_g}\right)\right]
%\end{equation}
\begin{equation}\label{eta}
\eta=-\sum_fg_f\int\frac{d^3p}{(2\pi)^3} \frac{|\bm{p}|^4}{30 \epsilon_f^2}  \bigg\{ \left(\frac{\partial f^0}{\partial \epsilon_f}\right)\left(A_1 e^{\chi_1}+A_2 e^{\chi_2}\right)+\left(\frac{\partial \overline{f^0}}{\partial \epsilon_f}\right)\left(\overline{A_1}e^{\overline{\chi_1}}+\overline{A_2}e^{\overline{\chi_2}}\right)\bigg\}-2g_g\int\frac{d^3p}{(2\pi)^3}\frac{|\bm{p}|^4}{30\epsilon_g^2}\tau_g\left(\frac{\partial f^0_g}{\partial \epsilon_g}\right).
\end{equation}
To obtain the final results, we have used the following identity:
\begin{align}
    p^ip^jp^kp^l\longrightarrow \frac{1}{15}\lvert\bm{p}\rvert^4(\delta^{ij}\delta^{kl}+\delta^{ik}\delta^{jl}+\delta^{il}\delta^{jk}).
\end{align}
The steps arriving at Eq.\eqref{eta} have been given in Appendix \ref{AppendixD1}.
In the following subsection, we show that the general expression obtained in Eq.\eqref{eta} (with $\mu=0$) reduces to the constant field results present in the literature under necessary limits on the time dependent fields of our analysis.\\

\subsubsection{{\it Zero field result:}} The expression of Eq.\eqref{eta} can be reduced to the zero field result in two ways. Recall that in our analysis the form of the magnetic field is $B = B_0 \exp[-t/\tau_B]$. We can set the magnetic field to be zero by either setting the amplitude, $B_0$ to be zero or by setting, $t \rightarrow \infty$.  In $\eta$, the parameter time ($t$), arises in the factor $A_1 e^{\chi_1}+A_2 e^{\chi_2}$. The factor $A_1$ is given by, $\int e^{-\chi_1} dt$, with $\chi_1 = -(t/\tau_R) -i(qB_0 /\epsilon_f) \int e^{-t/\tau_B} dt$. Taking the limit $t \rightarrow \infty$, we see that,
\begin{equation}
   \int e^{(t/\tau_R) +i(qB_0 /\epsilon_f) \int e^{(-t/\tau_B)} dt} dt \rightarrow \int e^{(\infty/\tau_R) +i(qB_0 /\epsilon_f) \int e^{(-\infty/\tau_B)} dt} dt = \int e^{(\infty/\tau_R)} dt,
\end{equation}
and similarly, 
\begin{equation}
  \lim_{t \rightarrow \infty} e^{\chi_1} = e^{-(\infty/\tau_R)}.  
\end{equation}
Hence we get,
\begin{equation}
   \lim_{t \rightarrow \infty} A_1 e^{\chi_1} = \frac{\int e^{(\infty/\tau_R)} dt}{e^{(\infty/\tau_R)}}=\frac{0}{0}. 
\end{equation}
Applying the L'Hospital's rule to solve the above equation we get,
\begin{equation}
 \lim_{t \rightarrow \infty} A_1 e^{\chi_1} =\lim_{t \rightarrow \infty} \frac{d(\int e^{(t/\tau_R)} dt)}{d(e^{(t/\tau_R)})}=\lim_{t \rightarrow \infty} \frac{e^{(t/\tau_R)}}{e^{(t/\tau_R)}/\tau_R} = \lim_{t \rightarrow \infty} \tau_R = \tau_R.   
\end{equation}
Similarly, we get, $$\lim_{t \rightarrow \infty} A_2 e^{\chi_2} = \tau_R.$$
Hence,
\begin{equation}
    \eta =-\bigg[\sum_f {g_f} \left(\frac{1}{60\pi^2}\right)\int d p\frac{p^6}{\epsilon_f^2} \left(\frac{\partial f^0}{\partial \epsilon_f}\right)(2\tau_R)+g_g\left(\frac{1}{30 \pi^2}\right)\int d p \frac{p^6 \tau_g}{\epsilon_g^2} \left(\frac{\partial f^g}{\partial \epsilon_g}\right)\bigg],
\end{equation}
which is what we observe for shear viscosity (with $\mu=0$) in literature\cite{Rath2021-as}. The result can also be obtained with the limits, $qB_0 \rightarrow 0$, as shown in the Appendix \ref{AppendixG}. \\

\subsubsection{{\it Constant field result:}}\label{shearconst} We have seen that the magnetic field tends to zero when $\tau_B \rightarrow 0$, and we can obtain the constant field limit with $\lim\limits_{\tau_B \rightarrow \infty} B_0 e^{-t/\tau_B}$. Hence, to obtain the shear viscosity in the constant field limit we can enforce, $\lim\limits_{\tau_B \rightarrow \infty} \eta$. Again, as the $\tau_B$ dependence is in the term $A_1 e^{\chi_1}+A_2 e^{\chi_2}$, we get (for $\mu=0$), 
\begin{equation}
  \lim_{\tau_B \rightarrow \infty} A_1 = \int e^{(1/\tau_R-i\omega_c)t} dt =   \frac{e^{(1/\tau_R-i\omega_c)t}}{(1/\tau_R-i\omega_c)},
\end{equation}
where we have used $\lim\limits_{\tau_B \rightarrow \infty} \chi_1 = (i\omega_c -1/\tau_R)t$ and $\lim\limits_{\tau_B \rightarrow \infty}\omega_c = qB_0/\epsilon_f$. Hence,
\begin{equation}
    A_1 e^{\chi_1} = \frac{e^{(1/\tau_R-i\omega_c)t}e^{-(1/\tau_R-i\omega_c)t}}{(1/\tau_R-i\omega_c)}=\frac{1}{(1/\tau_R-i\omega_c)}.
\end{equation}
Similarly for $A_2 e^{\chi_2}$, we get, 
\begin{equation}
    A_2 e^{\chi_2} = \frac{1}{(1/\tau_R+i\omega_c)}.
\end{equation}
Hence, 
\begin{equation}
    A_1 e^{\chi_1}+A_2 e^{\chi_2} =\frac{1}{(1/\tau_R-i\omega_c)}+ \frac{1}{(1/\tau_R+i\omega_c)}=\frac{2\tau_R}{(1+\omega_c^2 \tau^2_R)},
\end{equation}
Substituting this in the equation for $\eta$, we get the constant field result (with $\mu=0$) found in the literature, 
\begin{equation}
    \eta =-\left[\sum_f \left(\frac{{g_f}}{60 \pi^2}\right)\int d p \frac{p^6}{\epsilon_f^2} \left(\frac{\partial f^0}{\partial \epsilon_f}\right)\left(\frac{2\tau_R}{1+\omega_c^2 \tau^2_R}\right)+\left(\frac{g_g}{30 \pi^2}\right)\int d p \frac{p^6 \tau_g}{\epsilon_g^2} \left(\frac{\partial f^g}{\partial \epsilon_g}\right)\right].
\end{equation}
We have shown that the kinetic theory results found in the literature \cite{Rath2021-as} for the shear viscosity arise as special limiting cases of our result for a QGP system in a time dependent electromagnetic field.

 %%%%%%%%%%%%%%%%%%%%%%%%%%%%%%%%%%%%%%%%%%%%%%%%%%
\begin{figure*}
    \centering
    \includegraphics[width=0.495\textwidth]{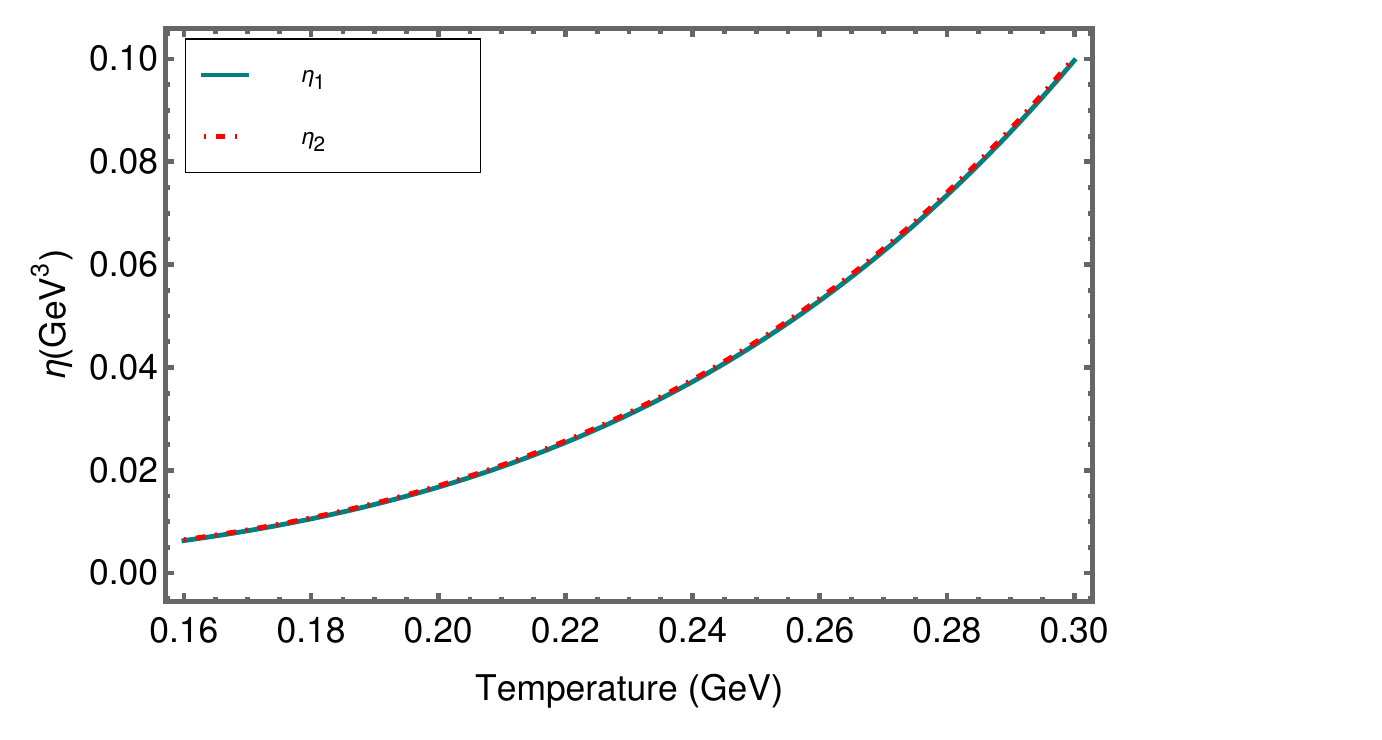}
    \includegraphics[width=0.465\textwidth]{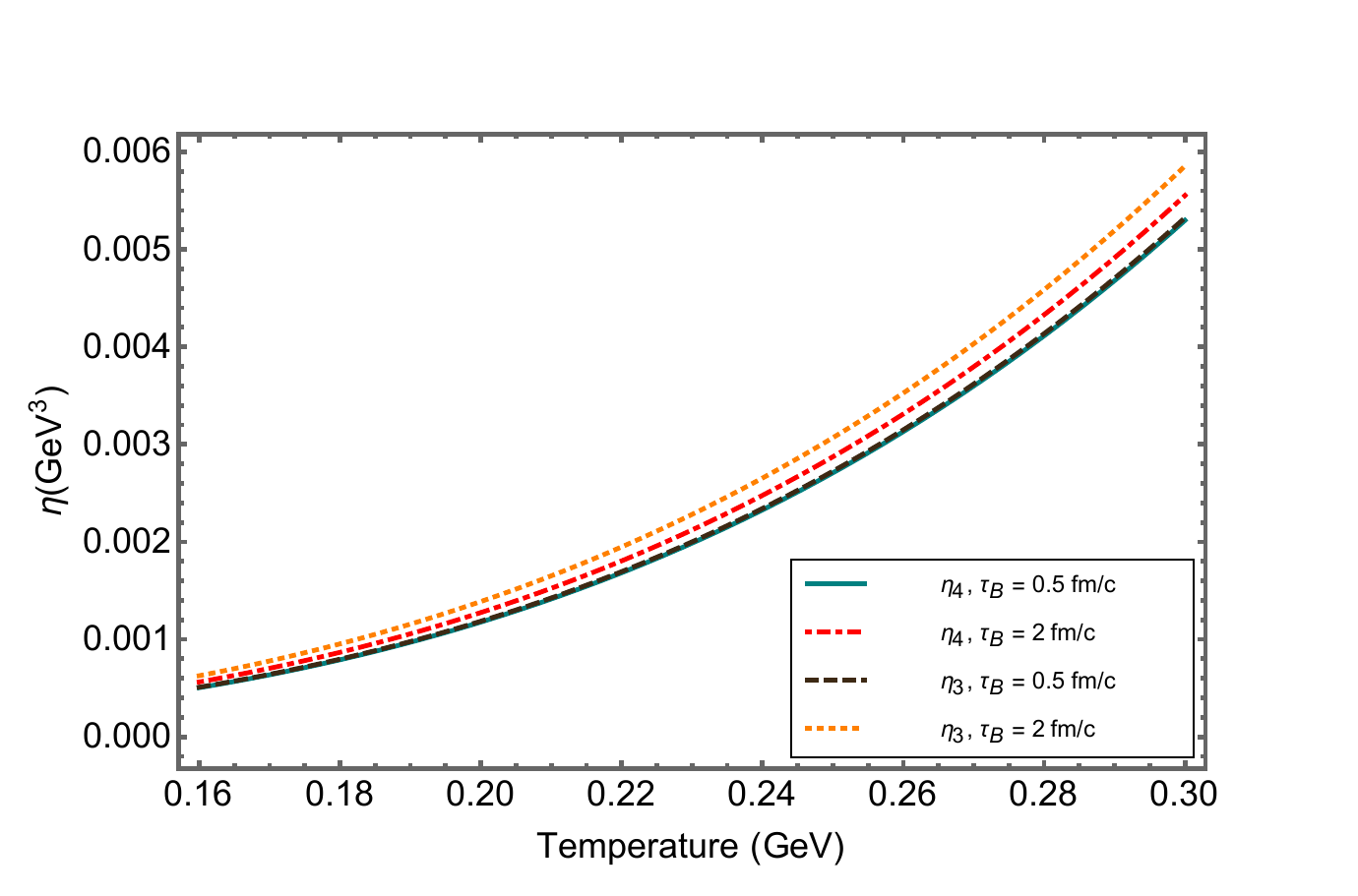}
    \hspace{-.5cm}
     \captionsetup{justification=raggedright, singlelinecheck=false, format=hang, labelsep=period}
    \caption{\small  Left panel: Temperature behavior of $\eta_{1,2}$ with $t=2$ fm/$c$, $qB_0=0.01$ GeV$^2$, and $\mu = 0.1$ GeV. Right panel: Temperature dependence of $\eta_{3,4}$ for two different values of magnetic field decay rate, $\tau_B$, with $qB_0=0.01$ GeV$^2$,  $t=2$ fm/$c$, and $\mu = 0.1$ GeV.}
\label{f6}
\end{figure*}
 %%%%%%%%%%%%%%%%%%%%%%%%%%%%%%%%%%%%%%%%%%%%%%%%%%
\subsection{Bulk viscosity}
The bulk viscosity is expressed in the dissipative part of the energy momentum tensor in terms of the gradient of the fluid flow. To obtain the bulk viscosity of the medium we compare Eq.\eqref{deltaT2} with Eq.\eqref{deltaT1} to obtain, 
%\begin{align}\label{41}
 %   \zeta = \sum_f \frac{g_f}{3} \int \frac{d^3 p}{(2 \pi)^3} \frac{p^2}{\epsilon_f} \bigg(\frac{\partial f^0}{\partial \epsilon_f}\bigg)A_f+\frac{g_g}{3} \int \frac{d^3 p}{(2 \pi)^3} \frac{p^2}{\epsilon_g} \bigg(\frac{\partial f^0}{\partial \epsilon_g}\bigg)A_g,
%\end{align}

\begin{align}\label{41}
    \zeta =\frac{1}{2} \sum_f \frac{g_f}{3} \int \frac{d^3 p}{(2 \pi)^3} \frac{p^2}{\epsilon_f}\left\{ \left(f_0(1-f_0\right)A_f+\overline{f_0}\left(1-\overline{f_0}\right)\overline{A_f}\right\}+\frac{g_g}{3} \int \frac{d^3 p}{(2 \pi)^3} \frac{p^2}{\epsilon_g} f^g_0\left(1+f^g_0\right)A_g,
\end{align}

where we have used $p^i p^j =\frac{p^2}{3} \delta^{ij}$, $A_f = \frac{\beta}{3}(A_1 e^{\chi_1} +A_2 e^{\chi_2})\left[ \frac{p^2}{\epsilon_f}-3\epsilon_f\frac{\partial P}{\partial \epsilon}\right]$, $\overline{A_f} =\frac{\beta}{3} \left(\overline{A_1} e^{\overline{\chi_1}} +\overline{A_2} e^{\overline{\chi_2}}\right)\left[\frac{p^2}{\epsilon_f} -3\epsilon_f \frac{\partial P}{\partial \epsilon_f} \right]$, and  $A_g = \tau_g\frac{\beta}{3}\left[\frac{p^2}{\epsilon_g}-3\epsilon_g \frac{\partial P}{\partial \epsilon_g}\right]$. 

The calculation of viscosity requires nonzero velocity gradient. But there exists different frames to define the fluid velocity $u^{\mu}$. For example, $u^{\mu}$ denotes the velocity of baryon number flow in the Eckart frame, whereas it denotes the velocity of energy flow in the Landau–Lifshitz frame. Therefore, the freedom to choose a specific frame creates arbitrariness. To avoid this arbitrariness, one needs the “condition of fit”, \textit{i.e.}, if one chooses the Landau–Lifshitz frame, then the condition of fit in the local rest frame demands the “$00$” component of the dissipative part of the energy–momentum tensor to be zero ($\Delta T^{00} = 0$). In order to satisfy this Landau–Lifshitz condition, the factors $A_f$, $\overline{A_f}$, and $A_g$ should be replaced as $A_f\rightarrow A_f'= A_f-c_f\epsilon_f$, $\overline{A_f}\rightarrow \overline{A_f'}= \overline{A_f}-\overline{c_f}\epsilon_f$, and $A_g\rightarrow A_g'= A_g-c_g\epsilon_g$. The Landau–Lifshitz conditions for $A_f$, $\overline{A_f}$, and $A_g$ are respectively given by\cite{PhysRevC.83.014906}, 
\begin{align}
    &\sum_f \frac{g_f}{3} \int \frac{d^3 p}{(2 \pi)^3} \epsilon_f\,f_0(1-f_0)(A_f-c_f \epsilon_f) = 0\label{43},\\
    &\sum_f \frac{g_f}{3} \int \frac{d^3 p}{(2 \pi)^3} \epsilon_f\,\overline{f_0}(1-\overline{f_0})(\overline{A_f}-\overline{c_f} \epsilon_f) = 0\label{43a},\\
    &\frac{g_g}{3} \int \frac{d^3 p}{(2 \pi)^3} \epsilon_g\,f^g_0(1+f^g_0)(A_g-c_g \epsilon_g) = 0.\label{44}
\end{align}
The quantities $c_f$, and $c_g$ are arbitrary constants and are associated with the particle and energy conservations for a thermal medium having asymmetry between the numbers of particles and antiparticles. These quantities can be obtained after substituting $A_f \rightarrow A^{'}_f$ , $\overline{A_f}\rightarrow \overline{A_f'}$, and $A_g \rightarrow A^{'}_g$, and solving for the arbitrary constants, $c_f$, $\overline{c_f}$, and $c_g$ from Eqs.[\ref{43}-\ref{44}]. Thus,  we obtain the final expression for the bulk viscosity,
%\begin{equation}
%    \zeta = \frac{1}{9} g_f \beta \int \frac{d^3 p}{(2 \pi)^3\epsilon_f^2}f^0(1-f^0)[p^2-3\epsilon_f^2 v_s^2]^2(A_1 e^{\chi_1} +A_2 e^{\chi_2})+\frac{1}{9} g_g \beta \int \frac{d^3 p}{(2 \pi)^3\epsilon_g^2}f_g^0(1-f^0_g)\tau_g[p^2-3\epsilon_g^2 v_s^2]^2
%\end{equation}
\begin{multline}\label{zeta}
    \zeta =\frac{\beta}{18}\sum_fg_f\int \frac{d^3 p}{(2 \pi)^3\epsilon_f^2}\left\{ f_0\left(1-f_0\right)\left(A_1e^{\chi_1}+A_2e^{\chi_2}\right)+\overline{f_0}\left(1-\overline{f_0}\right)\left(\overline{A_1}e^{\overline{\chi_1}}+\overline{A_2}e^{\overline{\chi_2}}\right)\right\}\left[p^2-3v_s^2\epsilon_f^2\right]^2+\\ \frac{\beta g_g}{9} \int \frac{d^3 p}{(2 \pi)^3\,\epsilon_g^2}  \tau_g\,f^g_0\left(1+f^g_0\right)\left[p^2-3v_s^2\epsilon_g^2\right]^2.
\end{multline}
The steps from Eq.\eqref{41} to Eq.\eqref{zeta} are detailed in Appendix \ref{AppendixF1}. We have used the following useful formulae for the pressure ($P$), entropy ($s$), the specific heat ($C_V$) and the velocity of sound squared ($v_s^2$) 
\begin{align}
    &P = T \int \frac{d^3 p}{(2 \pi)^3} f^{0}(1- f^{0}),\label{pressure}\\
    &s = \frac{1}{3T^2} \int \frac{d^3 p}{(2 \pi)^3}|\textbf{p}|^2 f^{0}(1- f^{0}),\label{entropy}\\
    &C_V = \frac{1}{T^2} \int \frac{d^3 p}{(2 \pi)^3}\epsilon_f^2 f^{0}(1- f^{0}),\label{cv}\\
    &v_s^2 = \frac{s}{C_V}\label{sound}.
\end{align}
In  what follows, we show that the general expression obtained in Eq.\eqref{zeta} reduces to the constant field results present in the literature under necessary limits on the time dependent fields of our analysis.\\ 

\subsubsection{{\it Zero field result:}} The expression Eq.\eqref{zeta} can be reduced to the zero field result in two ways. Recall that in our analysis, the form of the magnetic field is $B = B_0 \exp[-t/\tau_B]$. We can set the magnetic field to be zero by either setting the amplitude, $B_0$ to be zero or by setting, $t \rightarrow \infty$. In $\zeta$, the parameter time ($t$) arise in the factor $A_1 e^{\chi_1}+A_2 e^{\chi_2}$. The factor $A_1$ is given by, $\int e^{-\chi_1} dt$, with $\chi_1 = -(t/\tau_R) -i(qB_0 /\epsilon_f) \int e^{-t/\tau_B} dt$. Taking the limit $t \rightarrow \infty$, we see that,
\begin{equation}
   \int e^{(t/\tau_R) +i(qB_0 /\epsilon_f) \int e^{(-t/\tau_B)} dt} dt \rightarrow \int e^{(\infty/\tau_R) +i(qB_0 /\epsilon_f) \int e^{(-\infty/\tau_B)} dt} dt = \int e^{(\infty/\tau_R)} dt,
\end{equation}
and similarly, 
\begin{equation}
  \lim_{t \rightarrow \infty} e^{\chi_1} = e^{-(\infty/\tau_R)}.  
\end{equation}
Hence we get,
\begin{equation}
   \lim_{t \rightarrow \infty} A_1 e^{\chi_1} = \frac{\int e^{(\infty/\tau_R)} dt}{e^{(\infty/\tau_R)}}=\frac{0}{0}. 
\end{equation}
Applying the L'Hospital's rule to solve the above equation we get,
\begin{equation}
 \lim_{t \rightarrow \infty} A_1 e^{\chi_1} =\lim_{t \rightarrow \infty} \frac{d(\int e^{(t/\tau_R)} dt)}{d(e^{(t/\tau_R)})}=\lim_{t \rightarrow \infty} \frac{e^{(t/\tau_R)}}{e^{(t/\tau_R)}/\tau_R} = \lim_{t \rightarrow \infty} \tau_R = \tau_R.   
\end{equation}
Similarly, we get, $$\lim_{t \rightarrow \infty} A_2 e^{\chi_2} = \tau_R.$$
Hence, 
\begin{equation}
    \zeta = \frac{\beta}{36 \pi^2} \sum_f g_f \int dp \frac{p^2}{\epsilon_f^2}f^0(1-f^0)[p^2-3\epsilon_f^2 v_s^2]^2(2\tau_R)+\frac{g_g \beta}{18 \pi^2} \int d p \frac{p^2}{\epsilon_g^2}f_g^0(1+f^0_g)\tau_g[p^2-3\epsilon_g^2 v_s^2]^2,
\end{equation}
which is what we observe in literature \cite{Rath2021-as}. The result can also be obtained with the limits, $qB_0 \rightarrow 0$, as shown in the Appendix \ref{AppendixG}.

\subsubsection{{\it Constant field result:}}\label{bulkconst} The magnetic field tends to zero when $\tau_B \rightarrow 0$, and we can obtain the constant field limit with $\lim\limits_{\tau_B \rightarrow \infty} B_0 e^{-t/\tau_B}$. Hence, to obtain the bulk viscosity in the constant field limit, we can enforce, $\lim\limits_{\tau_B \rightarrow \infty} \zeta$. Again, as the $\tau_B$ dependence is in the term $A_1 e^{\chi_1}+A_2 e^{\chi_2}$, we get, 
\begin{equation}
  \lim_{\tau_B \rightarrow \infty} A_1 = \int e^{(1/\tau_R-i\omega_c)t} dt =   \frac{e^{(1/\tau_R-i\omega_c)t}}{(1/\tau_R-i\omega_c)},
\end{equation}
where we have used, $\lim\limits_{\tau_B \rightarrow \infty} \chi_1 = (i\omega_c -1/\tau_R)t$, and $\lim\limits_{\tau_B \rightarrow \infty}\omega_c = qB_0/\epsilon_f$. Hence,
\begin{equation}
    A_1 e^{\chi_1} = \frac{e^{(1/\tau_R-i\omega_c)t}e^{-(1/\tau_R-i\omega_c)t}}{(1/\tau_R-i\omega_c)}=\frac{1}{(1/\tau_R-i\omega_c)}.
\end{equation}
Similarly for $A_2 e^{\chi_2}$, we get, 
\begin{equation}
    A_2 e^{\chi_2} = \frac{1}{(1/\tau_R+i\omega_c)}.
\end{equation}
Hence, 
\begin{equation}
    A_1 e^{\chi_1}+A_2 e^{\chi_2} =\frac{1}{(1/\tau_R-i\omega_c)}+ \frac{1}{(1/\tau_R+i\omega_c)}=\frac{2\tau_R}{(1+\omega_c^2 \tau^2_R)},
\end{equation}
Substituting this in the equation for $\zeta$, we get the constant field result found in the literature, 
\begin{equation}
     \zeta = \frac{\beta}{36 \pi^2} \sum_f g_f \int dp \frac{p^2}{\epsilon_f^2}f^0(1-f^0)[p^2-3\epsilon_f^2 v_s^2]^2\bigg[\frac{2\tau_R}{(1+\omega_c^2 \tau^2_R)}\bigg]+\frac{g_g \beta}{18} \int d p \frac{p^2}{\epsilon_g^2}f_g^0(1+f^0_g)\tau_g[p^2-3\epsilon_g^2 v_s^2]^2.
\end{equation}
We have shown that the kinetic theory results found in the literature \cite{Rath2021-as} for the bulk viscosity arises as special limiting cases of our result for a QGP system in a time dependent electromagnetic field.

\subsection{Momentum transport coefficients in a general configuration
of magnetic field}
 In this section, we deal with the general decomposition of shear viscosity coefficients in the presence of an arbitrary time dependent background magnetic field. The time-independent case is well known with $\eta$ breaking up into five coefficients, and $\zeta$ into two. The starting point is to write the infinitesimal change in the distribution function of the charged particles in  a suitable tensor basis\cite{pitaevskii2012physical,Tuchin:2014hza},
\begin{equation}\label{50}
    \delta f = \sum_{l=0}^4 C_l Y_{mn}^l v_m v_n,
\end{equation}
where the coefficients are, 
\begin{align}
    Y_{ij}^0 =&\,(3b_i b_j -\delta_{ij})(b_k b_l V_{kl} -\frac{1}{3}\cdot \bm{V}) ,\\
    Y_{ij}^1 = & \,2V_{ij} +(3b_i b_j -\delta_{ij})\cdot \bm{V} +\delta_{ij}b_k b_l V_{kl}\nonumber\\
    &-2V_{ik}b_k b_j-2V_{jk}b_k b_i+b_i b_j V_{kl}b_k b_l,\\
    Y_{ij}^2 =& \,2V_{ik}b_k b_j +2V_{jk}b_k b_i -4b_i b_j V_{kl}b_k b_l ,\\
    Y_{ij}^3 =& \,V_{ik}b_{jk} +V_{jk}b_{ik}-V_{kl}b_{ik}b_j b_l-V_{kl}b_{jk}b_i b_l ,\\
    Y_{ij}^4 =& \,2V_{kl}b_{ik}b_j b_l+2V_{kl}b_{jk}b_i b_l ,\\
\end{align}
where $V_i$ and $b_i = \frac{B_i}{B}$, are respectively the fluid velocity, and the unit vector along the direction of the magnetic field, with $B_i$ being the components of the magnetic field \textbf{B}. $b_{ij} = \epsilon_{ijk}b_k$ and $V_{ij} = \frac{1}{2}(\frac{\partial v_i}{\partial x_j}+\frac{\partial v_j}{\partial x_i})$. The energy-momentum tensor in this basis can be written as, 
\begin{equation}\label{dtbasis}
    \Delta T^{ij} = \sum_{l=0}^4 \eta_l Y_{ij}^l,
\end{equation}
where $\eta_0$, $\eta_1$, $\eta_2$, $\eta_3$ and $\eta_4$ denote the five shear viscosity coefficients. The $\eta_0$ component is called the longitudinal viscosity as, $Y^0_{ij}b_i b_j \ne 0$ and the $\eta_{1,2,3,4}$ are called the transverse viscosities as, $Y^{1,2,3,4}_{ij}$ are transverse to $b_i b_j$. For the calculation of the viscosities, it is sufficient to take only spatial components of the nonequilibrium part of
the energy–momentum tensor. In the above tensor, we have excluded the bulk viscosity part to determine the shear viscosity coefficients. The change in the energy momentum tensor in terms of $\delta f$ is, 
\begin{equation}\label{52}
    \Delta T^{ij} = \sum _f g_f \int \frac{d^3 p}{(2 \pi)^3} \epsilon_f \,v^i v^j \delta f. 
\end{equation}
Substituting Eq.\eqref{50} in Eq.\eqref{52} we obtain the expression for $\Delta T^{ij}$,
\begin{equation}\label{dtij}
    \Delta T^{ij} = \frac{1}{15} \sum_f g_f \int \frac{d^3 p}{(2 \pi)^3} \epsilon_f v^4 (\delta_{ij} \delta_{mn}+\delta_{im}\delta_{jn}+\delta_{in}\delta_{jm}) \sum_{l=0}^4 C_l Y_{mn}^l,
\end{equation}
where again we have used the identity $v_iv_jv_mv_n\longrightarrow \frac{1}{15}\lvert\bm{v}\rvert^4(\delta_{ij}\delta_{mn}+\delta_{im}\delta_{jn}+\delta_{in}\delta_{jn}).$
  $\eta_0$ is the familiar zero magnetic field result:
\begin{equation}
    \eta_0 = \frac{\beta}{30 \pi^2} \sum_f g_f \int dp \frac{p^6}{\epsilon_f^2} \bigg[\tau_R f^0 (1-f^0)+\tau_R \bar{f}^0 (1-\bar{f}^0) \bigg].
\end{equation}
 %%%%%%%%%%%%%%%%%%%%%%%%%%%%%%%%%%%%%%%%%%%%%%%%%%
\begin{figure*}
    \centering
    \centering
    \hspace{-2.5cm}
    \includegraphics[width=0.485\textwidth]{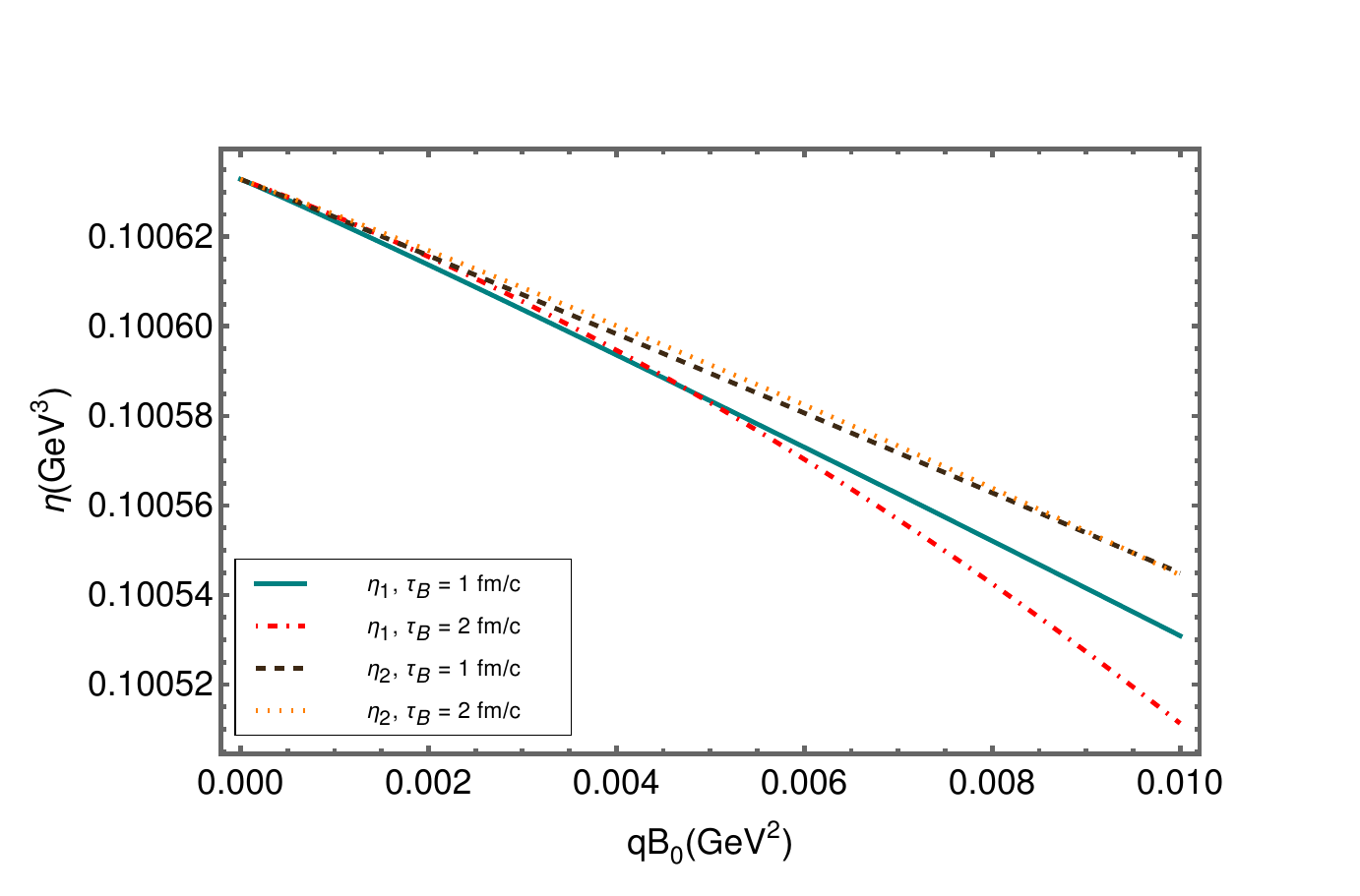}
    \hspace{-.5cm}
    \includegraphics[width=0.485\textwidth]{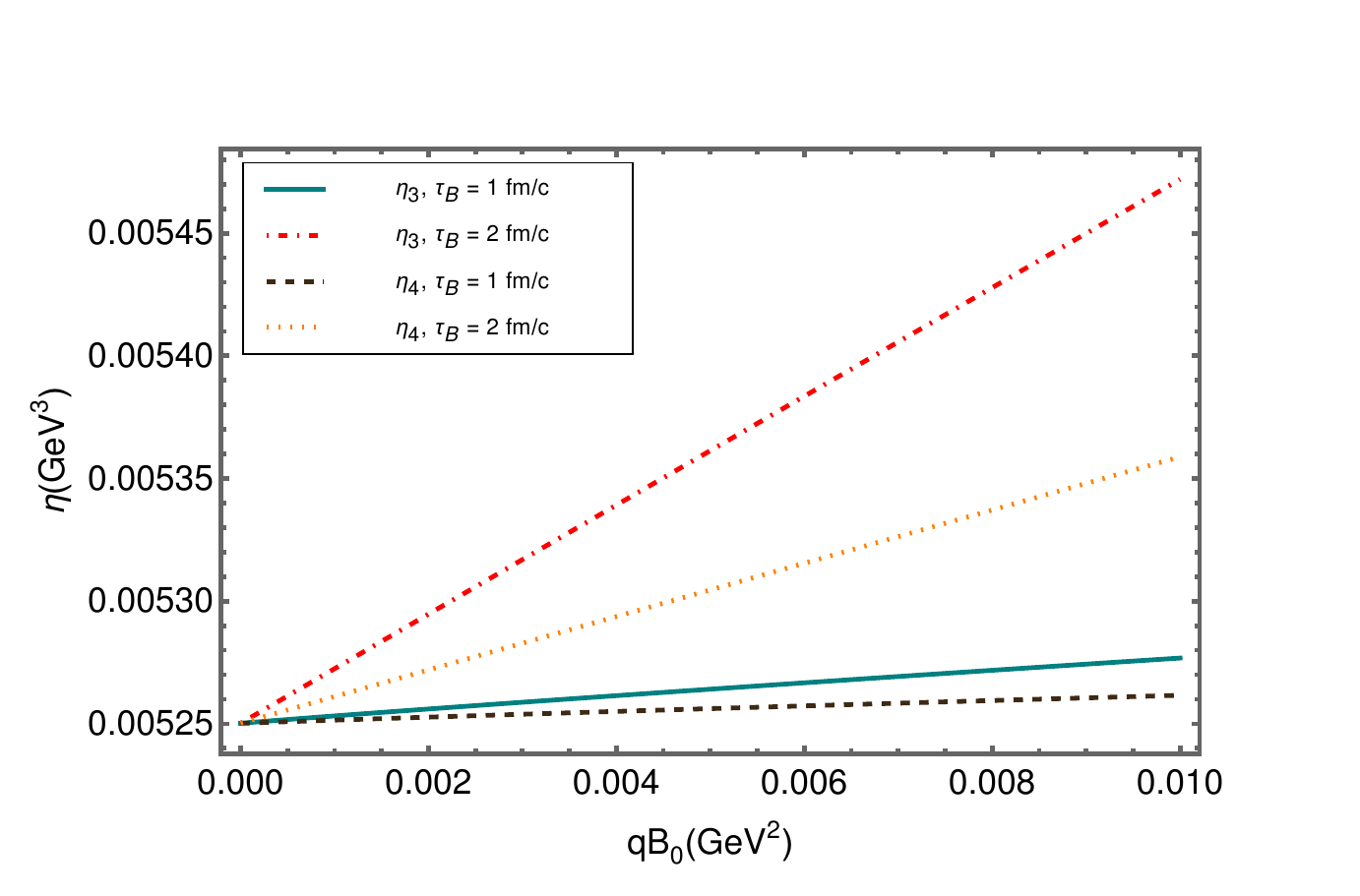}
    \hspace{-2.5cm}
    \captionsetup{justification=raggedright, singlelinecheck=false, format=hang, labelsep=colon}
    \caption{\small Left panel: Magnetic field ($qB_0$) dependence of $\eta_{1,2}$ for two  different values of the magnetic field decay rate, $\tau_B$, at $t=4$ fm/$c$, $T = 0.3$ GeV, and $\mu = 0.1$ GeV. Right panel: Magnetic field ($qB_0$) dependence of $\eta_{3,4}$ for two  different values of the magnetic field decay rate, $\tau_B$, at $t=4$ fm/$c$, $T = 0.3$ GeV, and $\mu = 0.1$ GeV.}
\label{f7}
\end{figure*}
 %%%%%%%%%%%%%%%%%%%%%%%%%%%%%%%%%%%%%%%%%%%%%%%%%%
 Comparing Eq.\eqref{dtij} and Eq.\eqref{dtbasis}, one arrives at the following expressions. These are derived in Appendix \ref{AppendixD}
\begin{align}
    &\eta_1 = \frac{2}{15} \sum_f g_f \int \frac{d^3 p}{(2 \pi)^3} \epsilon_f v^4 C_1,\label{etaa1}\\
    &\eta_2 = \frac{2}{15} \sum_f g_f \int \frac{d^3 p}{(2 \pi)^3} \epsilon_f v^4 C_2,\label{etaa2}\\
    &\eta_3 = \frac{2}{15} \sum_f g_f \int \frac{d^3 p}{(2 \pi)^3} \epsilon_f v^4 C_3,\label{etaa3}\\
    &\eta_4 = \frac{2}{15} \sum_f g_f \int \frac{d^3 p}{(2 \pi)^3} \epsilon_f v^4 C_4\label{etaa4}.
\end{align} 
To evaluate the coefficients $C_1, C_2, C_3$ and $C_4$ we need to solve the Boltzmann equation with $\delta f$ of the form in Eq.\eqref{50}. The Boltzmann equation reads,
\begin{equation}\label{begeneral}
    - \beta \epsilon_f V_{ij} v_i v_j f_0 (1-f_0)+qE^i\frac{\partial f^0}{\partial p^i}+\omega_c b_{ij}v_j \frac{\partial}{\partial v_i} \bigg(\sum_{l=0}^4 C_l Y_{mn}^l v_m v_n \bigg)+\sum_{l=0}^4 \dot{C}_l Y_{mn}^l v_m v_n = -\frac{\sum_{l=0}^4 C_l Y_{mn}^l v_m v_n}{\tau_R}.
\end{equation}
 Imposing the condition $\nabla\cdot V = 0$, and using the relations $V_{ij} b_i b_j = 0$, $b_{ij} v_i v_j = 0$, $b_i b_i = 1$, $b_{ij} b_i = 0$, and $b_{ij} b_j = 0$, the coefficients $C_i$ with $i=1,2,3,4$, are found to be,
\begin{align}
    &C_1 = \frac{\epsilon_f \beta f^0 (1-f^0)}{4}\bigg[\bigg(\int e^{-\xi_1}dt  \bigg)e^{\xi_1}+\bigg(\int e^{-\xi_2}dt  \bigg)e^{\xi_2}\bigg]\label{c1},\\
    &C_2 = \frac{\epsilon_f \beta f^0 (1-f^0)}{4}\bigg[\bigg(\int e^{-\xi_3}dt  \bigg)e^{\xi_3}+\bigg(\int e^{-\xi_4}dt\bigg)e^{\xi_4}\bigg]\label{c2},\\
     &C_3 = \frac{i\epsilon_f \beta f^0 (1-f^0)}{4}\bigg[\bigg(\int e^{-\xi_1}dt  \bigg)e^{\xi_1}-\bigg(\int e^{-\xi_2}dt\bigg)e^{\xi_2}\bigg]\label{c3},\\
     &C_4 = \frac{i\epsilon_f \beta f^0 (1-f^0)}{4}\bigg[\bigg(\int e^{-\xi_3}dt  \bigg)e^{\xi_3}-\bigg(\int e^{-\xi_4}dt\bigg)e^{\xi_4}\bigg],\label{c4}
\end{align}
where $\xi_{1,2,3,4}$ are given by,
\begin{align}
    &\xi_1 = -\frac{t}{\tau_R} -2i\int\frac{qB}{\epsilon_f}dt\label{xi1},\\
    &\xi_2 = -\frac{t}{\tau_R} +2i\int\frac{qB}{\epsilon_f}dt\label{xi2},\\
    &\xi_3 = -\frac{t}{\tau_R} -i\int\frac{qB}{\epsilon_f}dt\label{xi3},\\
    &\xi_4 = -\frac{t}{\tau_R} +i\int\frac{qB}{\epsilon_f}dt\label{xi4},
\end{align}
Substituting Eqs.[\ref{c1}-\ref{xi4}] in Eqs.[\ref{etaa1}-\ref{etaa4}] we get,
\begin{align}
    &\eta_1 = \frac{1}{60 \pi^2} \sum_f g_f \int dp \frac{p^6}{\epsilon_f^2}\beta f^0 (1-f^0)\bigg[e^{\xi_1}\int e^{-\xi_1}dt+e^{\xi_2} \int e^{-\xi_2}dt\bigg]\label{eta1},\\
    &\eta_2 = \frac{1}{60 \pi^2} \sum_f g_f \int dp \frac{p^6}{\epsilon_f^2}\beta f^0 (1-f^0)\bigg[e^{\xi_3}\int e^{-\xi_3}dt+e^{\xi_4} \int e^{-\xi_4}dt\bigg]\label{eta2},\\
    &\eta_3 = \frac{i}{60 \pi^2} \sum_f g_f \int dp \frac{p^6}{\epsilon_f^2}\beta f^0 (1-f^0)\bigg[e^{\xi_1}\int e^{-\xi_1}dt-e^{\xi_2} \int e^{-\xi_2}dt\bigg]\label{eta3},\\
    &\eta_4 = \frac{i}{60 \pi^2} \sum_f g_f \int dp \frac{p^6}{\epsilon_f^2}\beta f^0 (1-f^0)\bigg[e^{\xi_3}\int e^{-\xi_3}dt-e^{\xi_4} \int e^{-\xi_4}dt\bigg]\label{eta4}.
\end{align}
Eq.\eqref{begeneral}, followed by the following results have been derived in Appendix \ref{AppendixE}. The expressions above correspond to the shear viscosity of the charged particles in the presence of an arbitrary time-dependent magnetic field. As the gluons are electrically neutral, they are not affected by the external electromagnetic fields. Hence one can add the zero field gluon result to the above equations to obtain the complete expression for the shear viscosity, like in Eq.\eqref{eta}.
%%%%%%%%%%%%%%%%%%%%%%%%%%%%%%%%%%%%%%%%%%%%%%%%%%
\begin{figure*}
    \centering
    \centering
    \hspace{-2.5cm}
    \includegraphics[width=0.485\textwidth]{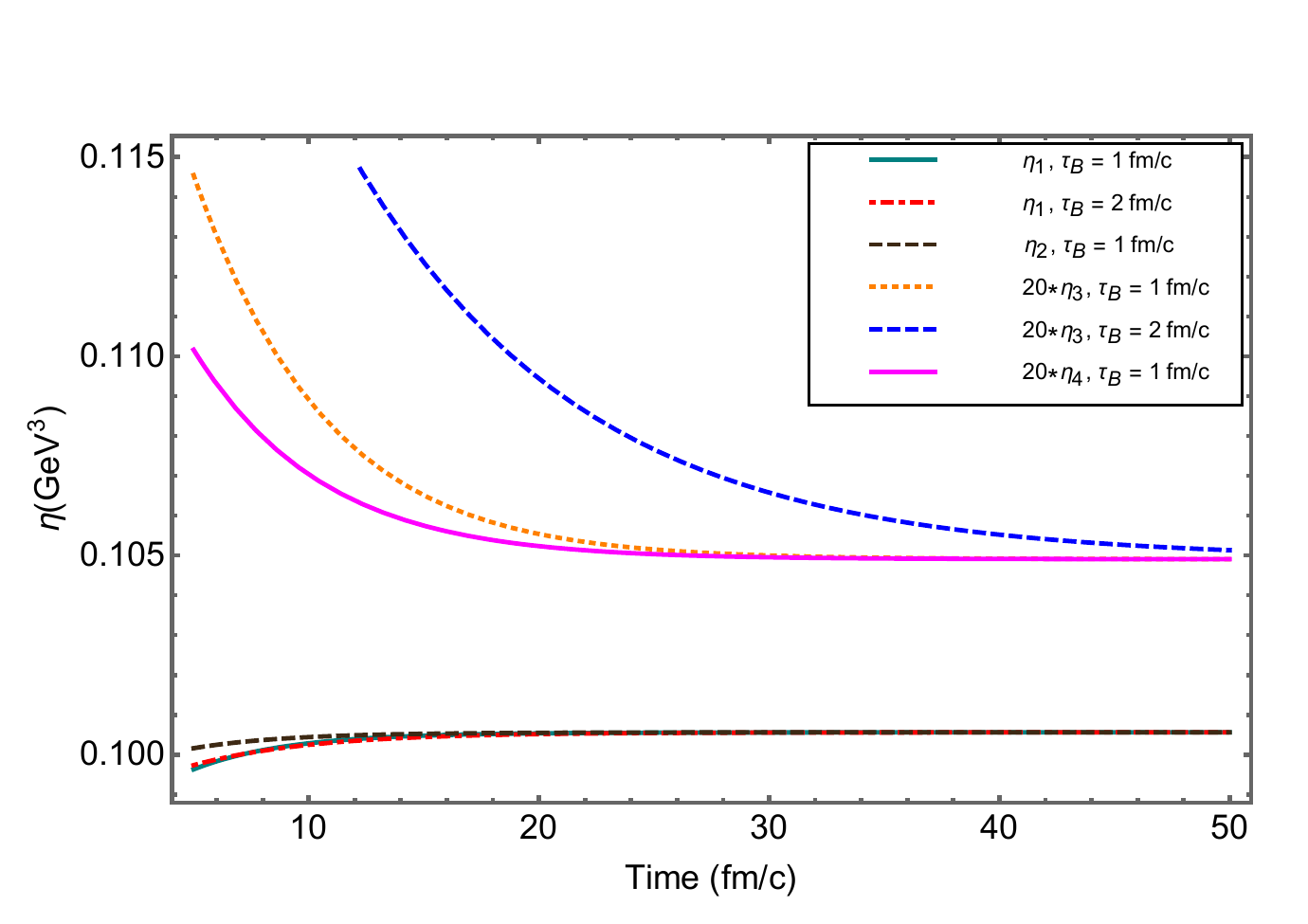}
    \hspace{-2.5cm}
    \captionsetup{justification=raggedright, singlelinecheck=false, format=hang, labelsep=period}
    \caption{\small Time dependence of $\eta_{1,2,3,4}$ at various values of the magnetic field decay rate, $\tau_B$, at $T = 0.3$ GeV, $qB_0 = 0.01$ GeV$^2$, and $\mu = 0.1$ GeV. }
\label{f8}
\end{figure*}
 %%%%%%%%%%%%%%%%%%%%%%%%%%%%%%%%%%%%%%%%%%%%%%%%%%

\section{Phenomenologically significant quantities}

\subsection{$\eta /s$ and $\zeta /s$}
We estimate the entropy density using equilibrium distribution functions as, 
\begin{equation}
    s = \frac{1}{3T^2} \int \frac{d^3 p}{(2 \pi)^3}|\textbf{p}|^2 f^{0}(1- f^{0}).
\end{equation}
 $\eta/s$ and $\zeta/s$ act as important quantities that capture the strength of the interaction of the system. The estimations from hydrodynamics and the KSS bound of the $\eta/s$ values have been crucial in understanding the QGP system as a strongly interacting matter and in establishing the importance of viscous effects in the hydrodynamical description of the QGP matter. From atomic and molecular physics we know that the shear viscosity has a minima and bulk viscosity exhibits a maxima near the vicinity of the phase transition. The estimation of the phase transition of the QGP medium is an ongoing and crucial open question in the heavy-ion collision experiments and the shear and bulk viscosities can act as crucial signals in their estimation.    

\subsection{Thermalization time}

The temperature of the QGP system created in heavy-ion collisions initially varies with space and time. The matter interact with each other and achieve thermal equilibrium, the time taken for which is called the thermalization time, $\tau_{th}$. Thermal equilibrium is vital for the implementation of hydrodynamics. There is debate on the thermalization time of the medium and recent results suggest that the unstable modes induced by the rapid expansion of the QGP fireball can affect the isotropization time and hence the thermalization time. The equation of state (EOS) is described for a system assuming isotropization and thermalization. Hence, the estimation of the thermalization time is crucially important in the study of the QGP system. With the shear viscosity over entropy density ratio $\eta/s$ characterizing the interaction strength, and temperature $T$ as the only scale, the time for the medium to thermalise is given by,
\begin{equation}
    \tau_{th} = 4 \pi \eta/Ts,
\end{equation}
The effect of the time dependent electromagnetic fields on the thermalization time has been studied through the shear viscosity of the system.

\section{Results and discussion}
The relaxation time in the R.H.S. of Eq.\eqref{BE} for quarks and gluons is given by\cite{Hosoya:NPB250'1985},
\begin{align}\label{taur}
\tau_R(T,\mu)&=\frac{1}{5.1\,T\alpha_{sq}(T,\mu)^2\log(1/\alpha_{sq}(T,\mu))[1+0.12(2N_f+1)]} ,   \\
\tau_g(T)&=\frac{1}{22.5\,T\alpha_{sg}(T)^2\log(1/\alpha_{sg}(T))[1+0.06N_f]},
\end{align}
where $\alpha_{sq}$, $\alpha_{sg}$ are the running couplings for quarks and gluons, respectively. For quarks, the expression is\cite{Ayala:PRD3'2018},
\begin{align}
    \alpha_{sq}[\Lambda^2(T,\mu),qB_0]=\frac{\alpha_s[\Lambda^2(T,\mu)]}{1+b_0\alpha_s(\Lambda^2(T,\mu))\ln\frac{\Lambda^2(T,\mu)}{\Lambda^2(T,\mu)+qB_0}},
\end{align}
with
\begin{equation}
  \alpha_s(\Lambda^2)=\frac{1}{b_0\ln\frac{\Lambda^2}{\Lambda^2_{\overline{MS}}}},  
\end{equation}
where $\Lambda(T,\mu)=2\pi \sqrt{T^2+\frac{\mu^2}{\pi^2}}$, $b_0=\frac{11N_c-2N_f}{2\pi}$. $\Lambda_{\overline{MS}}$ is set at 0.176 GeV. For gluons, $\alpha_{sg}$ is obtained simply by setting $\mu=0$ so that $\Lambda(T)=2\pi T$.
%We initiate our discussion of the viscous property of the QGP medium in response to time dependent electromagnetic field with the study of the Shear and Bulk viscosities. The Shear viscosity represents the momentum transport of the medium in the transverse medium and is quantified by the quantity $\eta$. 

In our analysis, the shear viscosity depends on the temperature, chemical potential and the time, decay rate,  amplitude/strength of the magnetic field at which the viscosity is calculated, \textit{i.e.}, $\eta\equiv\eta (T, \mu, t, \tau_B, qB_0)$. The form of $\eta$ is given in Eq.\eqref{eta}, where $A_1, A_2,\chi_1$ and $\chi_2$ are time dependent quantities that arise due to the nature of the magnetic field. In Fig.~\ref{f1}, we plot the shear viscosity as a function of temperature with the time fixed at $2$ fm/$c$ and chemical potential taken to be zero for various values of the strength of the magnetic field, $qB_0$, (left panel) and for various values of the decay rate of the magnetic field, $\tau_B$(right panel). The effect of $qB_0$ is seen to be small compared to the temperature. This is so because we are working in the weak field limit where the temperature dominates over the magnetic field. Hence, the effect of the magnetic field is more visible in the lower temperature region and is seen in the inset graph. We observe that as the  magnitude of the magnetic field increases, it leads to a decrease in the shear viscosity value. In the right panel, we see the effect of the decay rate on the shear viscosity at $qB_0 =0.01$ GeV$^2$. A faster decaying magnetic field has a stronger impact on $\eta$.
 %%%%%%%%%%%%%%%%%%%%%%%%%%%%%%%%%%%%%%%%%%%%%%%%%%
\begin{figure*}
    \centering
    \centering
    \hspace{-2.5cm}
    \includegraphics[width=0.485\textwidth]{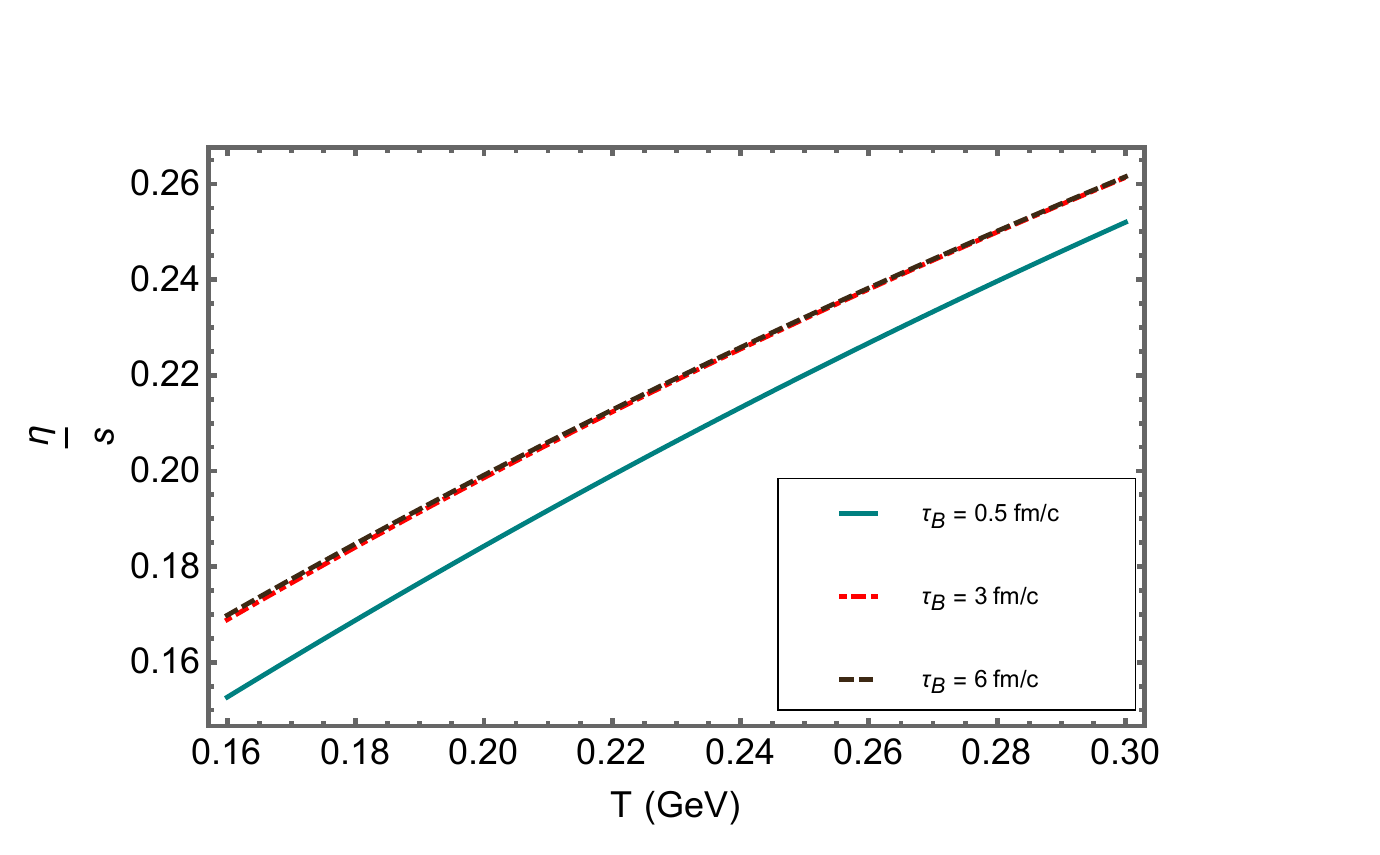}
    \hspace{-.5cm}
    \includegraphics[width=0.485\textwidth]{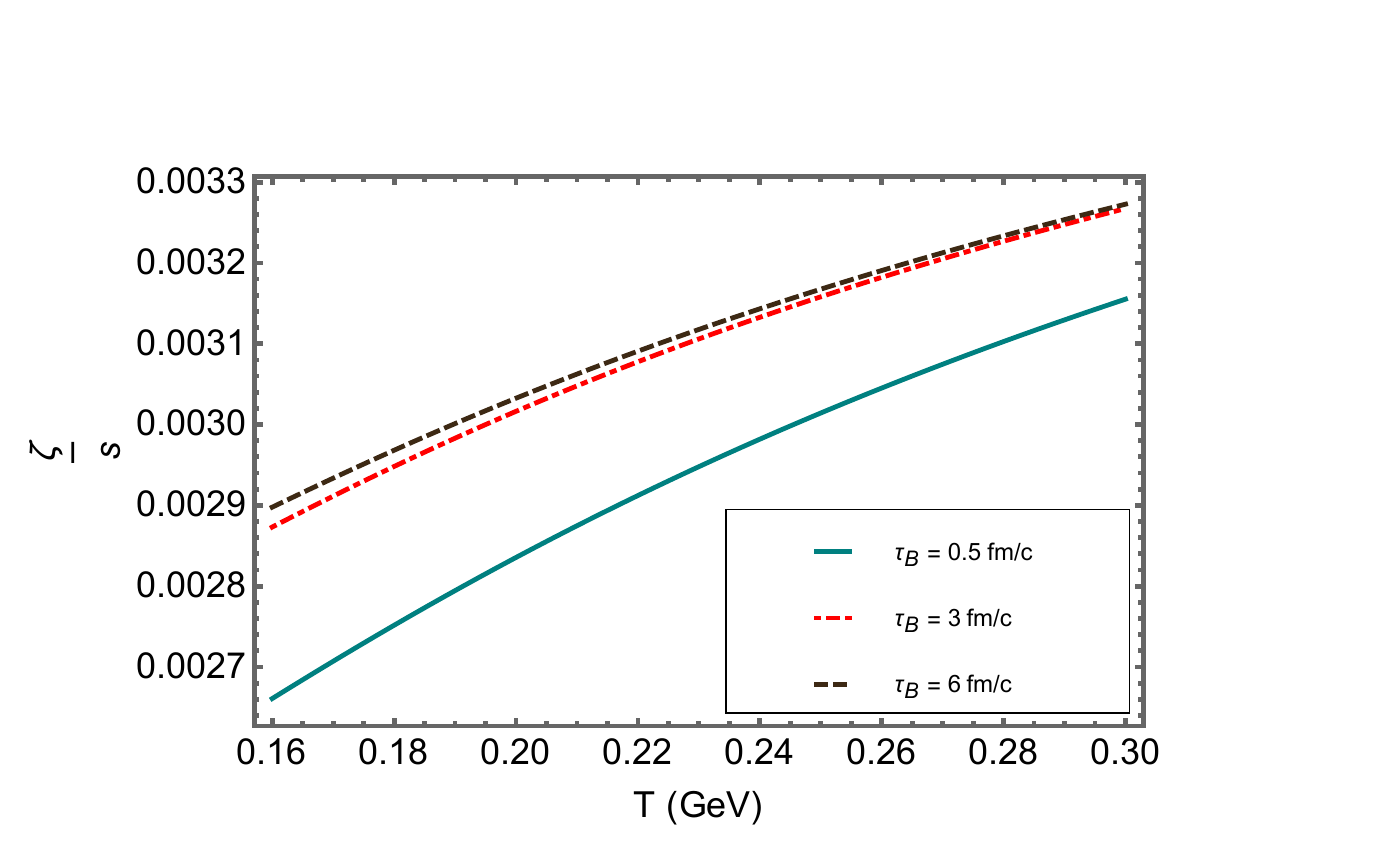}
    \hspace{-2.5cm}
    \captionsetup{justification=raggedright, singlelinecheck=false, format=hang, labelsep=period}
    \caption{\small Left panel: Temperature behavior of $\eta/s$ for various values of the magnetic field decay rate, $\tau_B$, with $t=1$ fm/$c$, $qB_0 = 0.02$ GeV$^2$ and $\mu = 0.1$ GeV. Right panel: Temperature dependence of $\zeta/s$ for various values of the magnetic field decay rate, $\tau_B$, with $t=1$ fm/$c$, $qB_0 = 0.02$ GeV$^2$ and $\mu = 0.1$ GeV.} 
\label{f9}
\end{figure*}
 %%%%%%%%%%%%%%%%%%%%%%%%%%%%%%%%%%%%%%%%%%%%%%%%%%

In Fig.~\ref{f2}, we look at the effects of time ($t$) and chemical potential ($\mu$) on the shear viscosity. The time dependency of the shear viscosity enters through the magnetic field which is parametrized by $\tau_B$, which tells us how fast the magnetic field decays with time. On the left panel, we plot the shear viscosity as a function of time for various values of $\tau_B$ at $qB_0 = 0.01$ GeV$^2$ and $T=0.3$ GeV. The behavior of the shear viscosity depends on the time scale under consideration. In the early time regime, a larger $\tau_B$ increases the shear viscosity while in the later time regime the faster decaying magnetic field, \textit{i.e.}, a lower values of $\tau_B$, increases the shear viscosity. The time behavior of the shear viscosity is dependent on the decay rate ($\tau_B$) and the relaxation time ($\tau_R$) and the interplay between them as can be seen in the time dependent part of the shear viscosity formula,
\begin{align}\label{timed}
&\chi_1 = -(t/\tau_R) -i(qB_0 /\epsilon_f) \int e^{-t/\tau_B} dt, &&\chi_2 = -(t/\tau_R) +i(qB_0 /\epsilon_f) \int e^{-t/\tau_B} dt.
\end{align}
These quantities influence $A_1e^{\chi_1}+A_2 e^{\chi_2}$ and hence, the shear and bulk viscosities. When we fix the temperature, we fix the relaxation time, $\tau_R$, and hence whether the $\tau_R$ term dominates or the $\tau_B qB_0$ term dominates in \eqref{timed}, depends on the time scale under consideration. The point of crossover depends on the magnitude of the magnetic field and the temperature fixed for consideration. In the right panel of Fig.~\eqref{f2}, we look at the dependency of the shear viscosity on the chemical potential. We plot for two different temperatures, $T=0.2, 0.3$ GeV at $t=2$ fm/$c$ and $\tau_B = 4$ fm/$c$. It is observed that the shear viscosity significantly increases with chemical potential. Also, the gap between the $\eta$ curves at two different temperatures  increases with chemical potential. 
 %%%%%%%%%%%%%%%%%%%%%%%%%%%%%%%%%%%%%%%%%%%%%%%%%%
\begin{figure*}
    \centering
    \centering
    \hspace{-2.5cm}
    \includegraphics[width=0.485\textwidth]{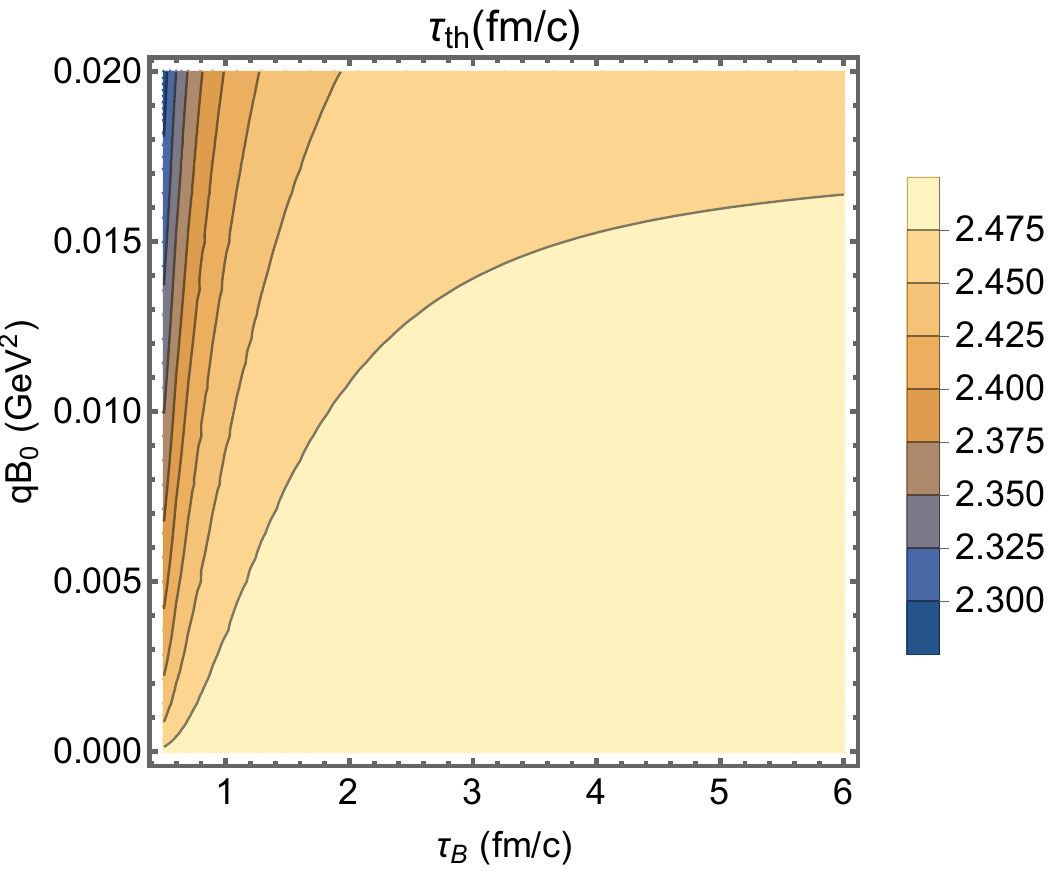}
    \hspace{-2.5cm}
    \captionsetup{justification=raggedright, singlelinecheck=false, format=hang, labelsep=period}
    \caption{\small  Contour graph of the thermalization time, $\tau_{th}$ as a function of the decay rate, $\tau_B$ and the amplitude of the magnetic field, $qB_0$, at $T=0.2$ GeV, $t=1$ fm/$c$ and $\mu = 0.1$ GeV.}
\label{f10}
\end{figure*}
 %%%%%%%%%%%%%%%%%%%%%%%%%%%%%%%%%%%%%%%%%%%%%%%%%%
The interdependency and the complex relationship between the magnitude and the decay rate of the magnetic field on the shear viscosity are captured in the contour graphs of the left panel of Fig.~\ref{f3}. We see that an increase in the decay rate of the magnetic field increases the shear viscosity. We observe that for the most part, the strength of the magnetic field plays a more important role than the decay rate on the behavior of the shear viscosity and the impact of $\tau_B$ increases with  $qB_0$. In the right panel of Fig.~\ref{f3}, we observe the dependency of the bulk viscosity on $qB_0$ and $\tau_B$ of the magnetic field. For the case of bulk viscosity, we observe that the $\tau_B$ has a stronger impact on $\zeta$, as compared to $\eta$. The bulk viscosity increases with the increase in $\tau_B$ and the impact of the decay rate increases with the strength of the magnetic field.    

 The bulk viscosity represents the viscous resistance to the change in the volume of the medium and is quantified by the quantity $\zeta$. In our analysis, the bulk viscosity depends on the temperature, chemical potential, time, decay rate, and amplitude of the magnetic field at which the viscosity is calculated, \textit{i.e.}, $\zeta\equiv\zeta (T, \mu, t, \tau_B, qB_0)$. The form of $\zeta$ is given in Eq.\eqref{zeta}, where $A_1, A_2,\chi_1$ and $\chi_2$ are time dependent quantities that arise due to the time dependent magnetic field. In Fig.~\ref{f4}, the plot of the bulk viscosity as a function of temperature with the time fixed at $2$ fm/$c$ and chemical potential to be zero, for various values of the strength of the magnetic field, $qB_0$ (left panel) and for various values of the decay rate of the magnetic field, $\tau_B$ (right panel), is examined. On the left panel, the effect of $qB_0$ is seen to be small compared to the temperature. The effect of the magnetic field is more visible in the lower temperature region and is seen in the inset graph. We observe that the increase in the magnitude of the magnetic field decreases the bulk viscosity. In the right panel, we see the effect of the decay rate on the bulk viscosity at $qB_0 = 0.01$ GeV$^2$. A faster decaying magnetic field has a stronger impact, and decreases $\zeta$.

The effects of time ($t$) and chemical potential ($\mu$) on the bulk viscosity are explored in Fig.~\ref{f5}. The time dependency of the bulk viscosity enters through the magnetic field which is parametrized by $\tau_B$. On the left panel, we plot the bulk viscosity as a function of time for various values of $\tau_B$ at $qB_0 = 0.01$ GeV$^2$ and $T=0.3$ GeV. The behavior of the bulk viscosity depends on the time under consideration. In the early time regime, a larger $\tau_B$ increases the bulk viscosity while in the later time regime the faster decaying magnetic field, \textit{i.e.}, a lower values of $\tau_B$, increases the bulk viscosity. The time behavior of the bulk viscosity is dependent on the decay rate ($\tau_B$) and the relaxation time ($\tau_R$) and the interplay between them. Similar to the analysis of the time behavior nature of the shear viscosity, and as can be seen in Eq.\eqref{timed}, the point of crossover depends on the magnitude of the magnetic field and the temperature fixed for consideration. In the right panel of Fig.~\ref{f5}, we look at the dependency of the bulk viscosity on the chemical potential. We plot for two different temperatures, $T=0.2, 0.3$ GeV at $t=2$ fm/$c$ and $\tau_B = 4$ fm/$c$. It is observed that the bulk viscosity significantly increases with chemical potential. Also, the gap between the $\zeta$ curves at two different temperatures  increases with chemical potential.  

The results of the decomposition of the shear viscosity, along and transverse to the direction of the magnetic field is shown in Fig.~\ref{f6}. On the left panel, the temperature behavior of $\eta_1$ and $\eta_2$ is explored and it is observed that they increase with temperature. On the right panel, we study the temperature dependency of $\eta_3$ and $\eta_4$ for various values of $\tau_B$. It is observed that $\eta_3$ is larger than $\eta_4$ and this difference is significant in the high temperature regime. The decay rate of the magnetic field, $\tau_B$, is seen to have a significant impact, and a faster decaying magnetic field is shown to decrease the viscosities. The effects of the temperature and decay rate are seen to be more pronounced for $\eta_3, \eta_4$ as compared to $\eta_1, \eta_2$. 

The magnetic field amplitude ($qB_0$) behavior of the decomposed viscosities are studied in Fig.~\ref{f7} for different values of the decay rate. On the left panel, we observe that $\eta_1,\eta_2$ decreases with an increase in $qB_0$. The increase in decay rate  is seen to decrease $\eta_1$. In the right panel, we study the $qB_0$ behavior of $\eta_3$ and $\eta_4$ for various values of $\tau_B$. We observe that the viscosities increase with an increase in $qB_0$ and the difference between $\eta_3$ and $\eta_4$ is enhanced in the high temperature region. The increase in the decay rate is seen to increase the viscosities, and the effect is more pronounced in $\eta_3$.

The time behavior of the decomposed shear viscosities are examined in Fig.~\ref{f8}. To keep the range of the results comparable we plot for $20 \times \eta_3$ and  $20 \times \eta_4$ along with $\eta_1$ and $\eta_2$. The low time behavior of $\eta_{1,2}$ differs as compared to $\eta_{3,4}$. The behavior of $\eta_{1,2}$ is seen to increase with an increase in time and both converge at large values. On the other hand $\eta_{3,4}$ decreases with time and the difference is also seen to be large compared to $\eta_{1,2}$.  $\tau_B$ is seen to increase $\eta_{3}$ and these two curves decay with time to an asymptotic value at large $t$. 

Now we look at some phenomenologically significant quantities. We begin with the results of $\eta/s$, which has been instrumental in establishing the QGP medium as a strongly interacting medium. In Fig.~\ref{f9}, on the left panel, we study the temperature behavior of $\eta/s$ for various values of $\tau_B$. The $\eta/s$ values obtained respects the K.S.S bound and is seen to increase with temperature. The $\eta/s$ ratio increases with an increase in the decay rate and this effect is seen to be more significant in the low temperature region. A faster decaying magnetic field leads to a smaller value of $\eta/s$. On the right panel, we study the temperature dependency of $\zeta/s$. The effect of the magnetic field and its decay rate is significant in the $\zeta/s$ results. We observe the $\zeta/s$ increases with an increase in the $\tau_B$ and its effect is more pronounced in the low temperature region. %The interaction strength of the medium, on which the $\eta/s$ and $\zeta/s$ values depend, in our analysis, is controlled by the relaxation time, $\tau_R$, and it is known that this decreases with temperature.

The behavior of thermalization time of the medium is examined in Fig.~\ref{f10} as a function of the decay rate and the strength of the magnetic field. The system is seen to thermalise faster for a larger amplitude of the magnetic field. The magnetic field is shown to have a significant effect, about \textcolor{blue}{$7\%$} for $qB_0$ values ranging from $0-0.02$ GeV$^2$, on the thermalization time. The estimates for the thermalization time agree with the perturbative QCD result of $2-3$ fm/$c$ \cite{BAIER200151,PhysRevC.71.064901,PhysRevLett.100.102301}.

\section{Conclusion and outlook}	
In conclusion, we have explored the momentum response of the QCD medium to time-varying electromagnetic fields. To that end, we have solved the relativistic Boltzmann equation within the relaxation time approximation. The shear and bulk viscosities are shown to depend on time, the amplitude and the decay rate of the magnetic field. The impact of the time dependent magnetic field is shown to be significant in low temperatures or in the regime of a fast-decaying magnetic field. The results are shown to perfectly reduce to the results in the literature for the zero and the constant field limits. The significance of the chemical potential and its interplay with the time dependent magnetic field is explored. The formalism was extended to the case of a general direction of the magnetic field and various components of the shear viscosities obtained. It was observed that the magnetic field and its time decaying nature have a relatively significant impact on the viscous coefficients. The behavior of all these quantities with temperature, time, decay rate and amplitude of the magnetic field has been explored. The shear viscosity to entropy density ratio and the bulk viscosity to entropy density ratio are seen to be critically dependent on the decay rate of the magnetic field where the fast decaying magnetic field result deviates significantly from a slowly decaying magnetic field, especially at lower temperatures. The thermalization time of the medium has also been estimated and the results are shown to be in the range of perturbative QCD results, and the decay rate has a significant impact on them.\\

The effects of time dependent magnetic fields are seen to be more prominent when the strength of the magnetic field is high, hence the development of the formalism for the strong magnetic field decaying with time, would be crucial and necessary for the complete understanding of the behavior of the QGP system.   A limitation of our present work is that the medium is considered to be static. An analysis with both the electromagnetic fields and the medium profile varying with time would lead to a more realistic result and we plan on taking up this problem in our future work.  The viscous coefficients are crucial in hydrodynamics, and hence observing their effects and their impact on the final state observables would be a fruitful endeavor. The challenging task of developing a resistive magnetohydrodynamics framework with time dependent electromagnetic fields would also be useful to the community.

\section{Acknowledgment}
G.K.K and D.D thank Sukanya Mitra for clarification of a mathematical doubt. G.K.K and D.D are thankful to the Indian Institute of Technology Bombay for the Institute postdoctoral fellowship. S. D. acknowledges the SERB Power Fellowship,
SPF/2022/000014 for the support on this work.

	\appendix
	\appendixpage
	\addappheadtotoc
	\begin{appendices}
		\renewcommand \thesubsection{\Alph{section}.\arabic{subsection}}
		\section{Derivation of $L$[Eq. \eqref{L}]}\label{AppendixA}
We use $L$ as a shorthand notation for the term $\tau_R\frac{p^{\mu}}{p^0}\frac{\partial  f_0}{\partial x^{\mu}}$. We begin with the Boltzmann equations for quarks, antiquarks, and gluons.
\begin{align}
 p^{\mu}\partial_{\mu}f&=\frac{-\epsilon_f}{\tau_R}\delta f \label{beq}  \\
  p^{\mu}\partial_{\mu}\bar{f}&=\frac{-\epsilon_f}{\tau_R}\delta \bar{f}\label{beaq} \\
   p^{\mu}\partial_{\mu}f_g&=\frac{-\epsilon_g}{\tau_g}\delta f_g \label{beg}
\end{align}
Rewriting Eq.\eqref{dem},
\begin{equation}
    \Delta T^{\mu\nu}=\int\frac{d^3p}{(2\pi)^3p^0}p^{\mu}p^{\nu}\delta f(x,p)
\end{equation}
Using Eq.s[\ref{beq}-\ref{beg}] in the expression for $\Delta T^{\mu\nu}$, we have, to first approximation,
\begin{equation}
    \Delta T^{\mu\nu}=-\int\frac{d^3p}{(2\pi)^3}\frac{p^{\mu}p^{\nu}}{(u\cdot p)}\left[\sum_fg_f\left(\frac{\tau_Rp^{\mu}\partial_{\mu} f_0+\tau_Rp^{\mu}\partial_{\mu} \bar{f}_0}{\epsilon_f}\right)+g_g\frac{\tau_gp^{\mu}\partial_{\mu}f^0_g}{\epsilon_g}\right]
\end{equation}
The Boltzmann equation, and hence, $\Delta T^{\mu\nu}$ is an iterative equation. In the first approximation, $f$ is simply replaced by $f^0$ so that we have 
\begin{equation}
    \Delta T^{\mu\nu}=-\int\frac{d^3p}{(2\pi)^3}p^{\mu}p^{\nu}\left[\sum_fg_f\left(\frac{L_q+L_{\bar{q}}}{\epsilon_f}\right)+g_g\frac{L_g}{\epsilon_g}\right]
\end{equation}
The projector $\Delta^{\mu\nu}\equiv g^{\mu\nu}-u^{\mu}u^{\nu}$ with $g^{\mu\nu}$ being the flat space metric, projects on to the space orthogonal to $u^{\mu}$. Then the derivative $\partial_{\mu}$ can be decomposed along and orthogonal to $u^{\mu}$ in the following way:
\begin{equation}
    \partial_{\mu}=\nabla_{\mu}+u_{\mu}D,
\end{equation}
\[ \text{where}\qquad \nabla_{\mu}\equiv \Delta^{\mu\nu}\partial_{\nu},\quad D\equiv u^{\mu}\partial_{\mu}\]
Then,
\begin{equation}
    \partial_{\mu}f^0=\nabla_{\mu}f^0+u_{\mu}Df^0
\end{equation}
For fermions, we have 
\begin{equation}
    f^0=\frac{1}{\exp{\beta(u^{\alpha}p_{\alpha})}+1}
\end{equation}
\begin{align}
  \nabla_{\mu}f^0&=\frac{f^0(1-f^0)}{T}\left[u^{\alpha}p_{\alpha}\frac{\nabla_{\mu}T}{T}-p_{\alpha}\nabla_{\mu}u^{\alpha}+T\nabla_{\mu}\left(\frac{\mu_f}{T}\right)\right] \label{nablaf} \\
  u_{\mu}Df^0&=\frac{f^0(1-f^0)}{T}\left[u_{\mu}u^{\alpha}p_{\alpha}\frac{DT}{T}-u_{\mu}p_{\alpha}Du^{\alpha}+Tu_{\mu}D\left(\frac{\mu_f}{T}\right)\right],\label{Df}
\end{align}
so that
\begin{equation*}
    \partial _{\mu}f^{0}=\beta f^0(1-f^0)\left[u^{\alpha}p_{\alpha}u_{\mu}\frac{DT}{T}+u^{\alpha}p_{\alpha}\frac{\nabla_{\mu}T}{T}-u_{\mu}p_{\alpha}Du^{\mu}-p_{\alpha}\nabla_{\mu}u^{\alpha}+T\partial_{\mu}\left(\frac{\mu_f}{T}\right)\right].
\end{equation*}
Similarly, for antiquarks and gluons, we have
\begin{align}
   \partial _{\mu}\bar{f}^{0}&=\beta \bar{f}^0(1-\bar{f}^0)\left[u^{\alpha}p_{\alpha}u_{\mu}\frac{DT}{T}+u^{\alpha}p_{\alpha}\frac{\nabla_{\mu}T}{T}-u_{\mu}p_{\alpha}Du^{\mu}-p_{\alpha}\nabla_{\mu}u^{\alpha}-T\partial_{\mu}\left(\frac{\mu_f}{T}\right)\right].\\
   \partial _{\mu}f_g^{0}&=\beta f_g^0(1-f_g^0)\left[u^{\alpha}p_{\alpha}u_{\mu}\frac{DT}{T}+u^{\alpha}p_{\alpha}\frac{\nabla_{\mu}T}{T}-u_{\mu}p_{\alpha}Du^{\mu}-p_{\alpha}\nabla_{\mu}u^{\alpha}\right].
\end{align}
From the energy-momentum conservation equation $\partial_{\mu}T^{\mu\nu}=0$, one can derive 
\begin{equation}\label{a14}
    \frac{DT}{T}=-\left(\frac{\partial P}{\partial \epsilon}\right)\nabla_{\alpha} u^{\alpha},\quad Du^{\alpha}=\frac{\nabla^{\alpha}P}{\epsilon+P}.
\end{equation}
On imposing the condition of fit $\Delta^{\mu\nu}=0$, only the spatial components of $\Delta^{\mu\nu}$ are non-zero. Further, we use the well-known relations
\begin{align}
    \partial_ku_l&=-\frac{1}{2}W_{kl}-\frac{1}{3}\delta_{kl}\partial_ju^j\\
    W_{kl}&=\partial_ku_l+\partial_lu_k-\frac{2}{3}\delta_{kl}\partial_ju^j\label{a16}
\end{align}
Using Eqs.[\ref{a14}-\ref{a16}], we finally arrive at
\begin{equation}
  \tau_R\frac{p^{\mu}}{p^0}\partial_{\mu}f^0\equiv L_q=  -\beta\tau_R f_0(1-f_0)\Bigg[\left\{\epsilon_f\left(\frac{\partial P}{\partial\epsilon}\right)-\frac{p^2}{3\epsilon_f}\right\}\partial_lu^l
  +p^l\left(\frac{\partial_l P}{\epsilon+P}-\frac{\partial_lT}{T}\right)-\frac{Tp^l}{\epsilon_f}\partial_l\left(\frac{\mu}{T}\right)-\frac{p^kp^l}{2\epsilon_f}W_{kl}\Bigg]
\end{equation}
Similarly, 
\begin{align}
    L_{\bar{q}}=-\beta\tau_R \bar{f}_0(1-\bar{f}_0)\Bigg[\left\{\epsilon_f\left(\frac{\partial P}{\partial\epsilon}\right)-\frac{p^2}{3\epsilon_f}\right\}\partial_lu^l
  +p^l\left(\frac{\partial_l P}{\epsilon+P}-\frac{\partial_lT}{T}\right)+\frac{Tp^l}{\epsilon_f}\partial_l\left(\frac{\mu}{T}\right)-\frac{p^kp^l}{2\epsilon_f}W_{kl}\Bigg]\\
  L_g=-\beta\tau_g f^0_g(1+f^0_g)\Bigg[\left\{\epsilon_g\left(\frac{\partial P}{\partial\epsilon}\right)-\frac{p^2}{3\epsilon_g}\right\}\partial_lu^l
  +p^l\left(\frac{\partial_l P}{\epsilon+P}-\frac{\partial_lT}{T}\right)-\frac{Tp^l}{\epsilon_g}\partial_l\left(\frac{\mu}{T}\right)-\frac{p^kp^l}{2\epsilon_g}W_{kl}\Bigg]
\end{align}

\section{Evaluation of $\Gamma_P$[Eq.\eqref{gammap}]}\label{AppendixB}
The particular solution of the differential equation
\begin{equation}\label{b1}
	\frac{d\bm{\Gamma}}{dt}=A\bm{\Gamma}+\bm{S}
\end{equation}
is evaluated via the method of variation of parameters. Following the standard procedure, it can be shown that the particular solution is given by
\begin{equation}\label{xp}
\bm{\Gamma_p}=x(t)\int x(t)^{-1}\bm{B},
\end{equation}
where $x(t)$ is the matrix composed of linearly independent solutions of the homogeneous differential equation  Eq.\eqref{deh}:
\begin{equation}
	x(t)=\begin{pmatrix}
		e^{\chi_1} & e^{\chi_2}\\
		-ie^{\chi_1} & ie^{\chi_2}
	\end{pmatrix}
\end{equation} 
The inverse of the matrix above is
\begin{equation}
\begingroup
\renewcommand*{\arraystretch}{1.5}
	x(t)^{-1}=\frac{1}{2ie^{(\chi_1+\chi_2)}}\begin{pmatrix}
		e^{\chi_1} & e^{\chi_2}\\
		-ie^{\chi_1} & ie^{\chi_2}
	\end{pmatrix}=\begin{pmatrix}
		\frac{1}{2} \, e^{-\chi_1} & \frac{i}{2} \, e^{-\chi_1}\\
		\frac{1}{2} \, e^{-\chi_2} & \frac{-i}{2}\, e^{-\chi_2}
	\end{pmatrix}
    \endgroup
\end{equation} 
Now, 
\begin{equation}
\begingroup
\renewcommand*{\arraystretch}{1.5}
	x(t)^{-1}\bm{B}=\begin{pmatrix}
		\frac{b_1}{2}  e^{-\chi_1} + \frac{ib_2}{2} \, e^{-\chi_1}\\
		\frac{b_1}{2} \, e^{-\chi_2} + \frac{-ib_2}{2}\, e^{-\chi_2}
	\end{pmatrix};
\endgroup
\end{equation} 
where
\begin{equation}
    b_1=-\tau_R q \dot{E_x}+q\tau_R\omega_cE_y+L_x\,,\quad b_2=-\tau_R q \dot{E_y}-q\tau_R\omega_cE_x+L_y
\end{equation}
Thus,  the first element of the column matrix $x(t)^{-1}\bm{B}$ is 
\begin{equation}
    \frac{-\tau_R q}{2}\int dt\,\dot{E_x}e^{-\chi_1}+\frac{q^2\tau_R}{2\epsilon_f}\int dt\, BE_y e^{-\chi_1}+\frac{L_x}{2}\int dt\,e^{-\chi_1}- \frac{i\tau_R q}{2}\int dt\,\dot{E_y}e^{-\chi_1}-\frac{iq^2\tau_R}{2\epsilon_f}\int dt\, BE_x e^{-\chi_1}+\frac{iL_y}{2}\int dt\,e^{-\chi_1}
\end{equation}
Let us look at the first integral, for which, we recall that $\chi_1=\int dt\left(1/\tau_R +\frac{iqB_0}{\epsilon_f}e^{-t/\tau_B}\right)=t/\tau_R-ibe^{-t/\tau_B}$, with $b=\frac{qB_0\tau_B}{\epsilon_f}$. By changing variables via $e^{-t/\tau_B}=u$, the first integral is written as
\begin{equation}
    I_1=\frac{-\tau_R q}{2}\frac{E_0\tau_B}{\tau_E}\int du\, u^{a-1}e^{-ibu},
\end{equation}
where $a=\tau_B(1/\tau_E-1/\tau_R)$. Similarly, for the remaining 5 integrals, we have
\begin{align}
 I_2&=\frac{-\tau_R\, \tau_B\, q^2B_0E_0}{2\epsilon_f}\int du\, u^{a_1-1}e^{-ibu}\\
 I_3&=\frac{-\tau_B\,L_x}{2}\int du\, u^{a_2-1}e^{-ibu}\\
 I_4&=\frac{-i\tau_R q}{2}\frac{E_0\tau_B}{\tau_E}\int du\, u^{a-1}e^{-ibu}\\
 I_5&=\frac{i\tau_R\, \tau_B\, q^2B_0E_0}{2\epsilon_f}\int du\, u^{a_1-1}e^{-ibu}\\
 I_6&=\frac{-\tau_B\,L_y}{2}\int du\, u^{a_2-1}e^{-ibu},
\end{align}
where
\begin{equation}
    a_1=\tau_B\left(\frac{1}{\tau_E}+\frac{1}{\tau_B}-\frac{1}{\tau_R}\right), \quad a_2=\frac{-\tau_B}{\tau_R}
\end{equation}
Thus, the first element of the matrix $x(t)^{-1}\bm{B}$ is $\sum\limits_{j=1}^6 I_j\equiv E_1 (\text{say})$. The second element, similarly, is 
$\sum\limits_{j=7}^{12} I_j\equiv E_2$, with
\begin{align}
 I_7&=\frac{-\tau_R\, \tau_B\, qE_0}{2\tau_E}\int du\, u^{a_1-1}e^{ibu}\\
 I_8&=\frac{i\tau_R\, \tau_B\, qE_0}{2\tau_E}\int du\, u^{a_1-1}e^{ibu}\\
 I_9&=\frac{-\tau_R\, \tau_B\, q^2B_0E_0}{2\epsilon_f}\int du\, u^{a_1-1}e^{ibu}\\
 I_{10}&=\frac{-i\tau_R\, \tau_B\, q^2B_0E_0}{2\epsilon_f}\int du\, u^{a_1-1}e^{ibu}\\
 I_{11}&=\frac{-\tau_B\,L_x}{2}\int du\, u^{a_2-1}e^{ibu}\\
 I_{12}&=\frac{i\tau_B\,L_y}{2}\int du\, u^{a_2-1}e^{ibu}
\end{align}
Finally, from Eq.\eqref{xp}, we have
\begin{equation}
    \begingroup
\renewcommand*{\arraystretch}{1.5}
	\bm{\Gamma_p}=\begin{pmatrix}
		e^{\chi_1}E_1 + e^{\chi_2}E_2\\
		-ie^{\chi_1}E_1 + ie^{\chi_2}E_2
	\end{pmatrix}\equiv \begin{pmatrix}
		\Gamma_{xp}\\
		\Gamma_{yp}
	\end{pmatrix}
    \endgroup
\end{equation}

\section{Derivation of $\Delta T^{ij}$ [Eq.\eqref{deltaT2}]}\label{AppendixC}
We recall that 
\begin{equation}
    \Delta T^{ij}=\int\frac{d^3p}{(2\pi)^3}p^{i}p^{j}\left[\sum_fg_f\left(\frac{\delta f_q+\delta \bar{f_q}}{\epsilon_f}\right)+g_g\frac{\delta f_g}{\epsilon_g}\right]
\end{equation}
Employing the Ansatz [Eq. \ref{ansatz}] for $\delta f$, we get
  \begin{multline}\label{tij}
    \Delta T^{ij}=-\sum_fg_f\int\frac{d^3p}{(2\pi)^3}\frac{p^{i}p^{j}}{\epsilon_f}\bigg[(\tau_R qE_xv_x+\tau_R qE_yv_y+v_x\Gamma_x+v_y\Gamma_y)\frac{\partial f_0}{\partial \epsilon_f}+\\(\tau_R\bar{q}E_xv_x+\tau_R\bar{q}E_yv_y+v_x\overline{\Gamma_x}+v_y\overline{\Gamma_y})\frac{\partial \bar{f_0}}{\partial \epsilon_f}+\,\text{gluonic contribution}\bigg]
\end{multline}  
Of all the terms in $\Gamma_x$, $\Gamma_y$, the relevant terms are only those which possess the velocity gradient structure, $i.e.$, $W_{kl}$. Such integrals are $I_3$, $I_6$, $I_{11}$ and $I_{12}$. So, effectively, Eq.s \eqref{gx} and \eqref{gy} reduce to
\begin{align}
\Gamma_x&=(I_3+I_6)e^{\chi_1}+(I_{11}+I_{12})e^{\chi_2}\,,\quad \Gamma_y=-i(I_3+I_6)e^{\chi_1}+i(I_{11}+I_{12})e^{\chi_2} \\[0.3em]
\overline{\Gamma_x}&=(\overline{I_3}+\overline{I_6})e^{\overline{\chi_1}}+(\overline{I_{11}}+\overline{I_{12}})e^{\overline{\chi_2}}\,,\quad \overline{\Gamma_y}=-i(\overline{I_3}+\overline{I_6})e^{\overline{\chi_1}}+i(\overline{I_{11}}+\overline{I_{12}})e^{\overline{\chi_2}},
\end{align}
for quarks and antiquarks respectively. Eq.\eqref{tij} now reads
\begin{equation}
   \Delta T^{ij}=-\sum_fg_f\int\frac{d^3p}{(2\pi)^3}\frac{p^{i}p^{j}}{\epsilon_f}\left(\frac{\partial f_0}{\partial \epsilon_f}\right)\left[\Gamma_xv_x+\Gamma_yv_y+\overline{\Gamma_x}v_x+\overline{\Gamma_y}v_y\right]  
\end{equation}
Now,
\begin{align}
    \Gamma_xv_x+\Gamma_yv_y&=\frac{1}{2}(L_xv_x+L_yv_y)(A_1e^{\chi_1}+A_2e^{\chi_2})+\, \text{cross terms of the type }\, L_xv_y,\,L_yv_x\\[0.3em]
\overline{\Gamma_x}v_x+\overline{\Gamma_y}v_y&=\frac{1}{2}(L_xv_x+L_yv_y)(\overline{A_1}e^{\overline{\chi_1}}+\overline{A_2}e^{\overline{\chi_2}})+\, \text{cross terms of the type }\,\overline{L_x}v_y,\,\overline{L_y}v_x,
\end{align}
where $A_1=\int dt\, e^{-\chi_1}\,,\quad A_2=\int dt \,e^{-\chi_2}$
Assuming $p_z=0$, this simplifies to
\begin{equation}
  L_xv_x+L_yv_y=\left[\epsilon_f\frac{\partial P}{\partial \epsilon}-\frac{p^2}{3\epsilon_f}\right]\partial_lu^l-\frac{p^kp^lW_{kl}}{2}  
\end{equation}
Finally, thus, we arrive at Eq.\eqref{deltaT2} 
\begin{multline}
 \Delta T^{ij}=-\frac{1}{2}\Bigg[\sum_fg_f\int\frac{d^3p}{(2\pi)^3}\frac{p^{i}p^{j}}{\epsilon_f}   \bigg\{ \left(\frac{\partial f^0}{\partial \epsilon_f}\right)\left(A_1 e^{\chi_1}+A_2 e^{\chi_2}\right)+\left(\frac{\partial \overline{f^0}}{\partial \epsilon_f}\right)\left(\overline{A_1}e^{\overline{\chi_1}}+\overline{A_2}e^{\overline{\chi_2}}\right)\bigg\}\times \bigg\{\left(\epsilon_f\frac{\partial P}{\partial \epsilon}-\frac{p^2}{3\epsilon_f}\right)\partial_lu^l-\\\frac{p^kp^lW_{kl}}{2\epsilon_f} \bigg\}\Bigg] +g_g\int\frac{d^3p}{(2\pi)^3}\frac{p^{i}p^{j}}{\epsilon_g}\tau_g\beta f_g^0(1+f_g^0)\times \left\{\left(\epsilon_g\frac{\partial P}{\partial \epsilon}-\frac{p^2}{3\epsilon_g}\right)\partial_lu^l-\frac{p^kp^lW_{kl}}{2\epsilon_g} +p^l\left(\frac{\partial_lP}{\epsilon+P}-\frac{\partial_lT}{T}\right)\right\}
\end{multline}

\section{Derivation of $\eta$ [Eq.\eqref{eta}]}\label{AppendixD1}
We look at the terms in $\Delta T^{ij}$ containing $W_{kl}$.
\begin{multline}\label{wkl}
 \Delta T^{ij}=-\frac{1}{2}\sum_fg_f\int\frac{d^3p}{(2\pi)^3}\frac{p^{i}p^{j}}{\epsilon_f}   \bigg\{ \left(\frac{\partial f^0}{\partial \epsilon_f}\right)\left(A_1 e^{\chi_1}+A_2 e^{\chi_2}\right)+\left(\frac{\partial \overline{f^0}}{\partial \epsilon_f}\right)\left(\overline{A_1}e^{\overline{\chi_1}}+\overline{A_2}e^{\overline{\chi_2}}\right)\bigg\}\times \\\left(\frac{-p^kp^lW_{kl}}{2\epsilon_f}\right) +g_g\int\frac{d^3p}{(2\pi)^3}\frac{p^{i}p^{j}}{\epsilon_g}\tau_g\beta f_g^0(1+f_g^0)\times \left(-\frac{p^kp^lW_{kl}}{2\epsilon_g}\right) 
\end{multline}
Using 
\begin{align}
    p^ip^jp^kp^l\longrightarrow \frac{1}{15}\lvert\bm{p}\rvert^4(\delta^{ij}\delta^{kl}+\delta^{ik}\delta^{jl}+\delta^{il}\delta^{jk}),
\end{align}
Eq.\eqref{wkl} becomes
\begin{multline}\label{wkl2}
 \Delta T^{ij}=\frac{1}{2}\sum_fg_f\int\frac{d^3p}{(2\pi)^3}\frac{W_{kl}}{2\epsilon_f^2} \frac{|\bm{p}|^4}{15}(\delta^{ij}\delta^{kl}+\delta^{ik}\delta^{jl}+\delta^{il}\delta^{jk})  \bigg\{ \left(\frac{\partial f^0}{\partial \epsilon_f}\right)\left(A_1 e^{\chi_1}+A_2 e^{\chi_2}\right)+\left(\frac{\partial \overline{f^0}}{\partial \epsilon_f}\right)\left(\overline{A_1}e^{\overline{\chi_1}}+\overline{A_2}e^{\overline{\chi_2}}\right)\bigg\}\times \\+g_g\int\frac{d^3p}{(2\pi)^3}\left(-\frac{W_{kl}}{2\epsilon_g^2}\right)\frac{|\bm{p}|^4}{15}(\delta^{ij}\delta^{kl}+\delta^{ik}\delta^{jl}+\delta^{il}\delta^{jk})\tau_g\beta f_g^0(1+f_g^0) 
\end{multline}
The delta functions contracted with $W_{kl}$ produce $W_{ll}$, $W_{ij}$, $W_{ji}$. The tensor $W_{ij}$ is defined so that $W_{ii}=0$, and $W_{ij}=W_{ji}$. Thus, finally,
\begin{multline}\label{wkl3}
\Delta T^{ij}=\frac{1}{2}\sum_fg_f\int\frac{d^3p}{(2\pi)^3}(2W^{ij}) \frac{|\bm{p}|^4}{30 \epsilon_f^2}  \bigg\{ \left(\frac{\partial f^0}{\partial \epsilon_f}\right)\left(A_1 e^{\chi_1}+A_2 e^{\chi_2}\right)+\left(\frac{\partial \overline{f^0}}{\partial \epsilon_f}\right)\left(\overline{A_1}e^{\overline{\chi_1}}+\overline{A_2}e^{\overline{\chi_2}}\right)\bigg\}\times \\+g_g\int\frac{d^3p}{(2\pi)^3}(2W^{ij})\frac{|\bm{p}|^4}{30\epsilon_g^2}\tau_g\beta f_g^0(1+f_g^0) 
\end{multline}

Comparing with Eq.\eqref{deltaT1}, one can read off the shear viscosity:
\begin{equation}
\eta=-\sum_fg_f\int\frac{d^3p}{(2\pi)^3} \frac{|\bm{p}|^4}{30 \epsilon_f^2}  \bigg\{ \left(\frac{\partial f^0}{\partial \epsilon_f}\right)\left(A_1 e^{\chi_1}+A_2 e^{\chi_2}\right)+\left(\frac{\partial \overline{f^0}}{\partial \epsilon_f}\right)\left(\overline{A_1}e^{\overline{\chi_1}}+\overline{A_2}e^{\overline{\chi_2}}\right)\bigg\}-2g_g\int\frac{d^3p}{(2\pi)^3}\frac{|\bm{p}|^4}{30\epsilon_g^2}\tau_g\left(\frac{\partial f^0_g}{\partial \epsilon_g}\right)
\end{equation}

\section{Alternative zero field limit derivation}\label{AppendixG}
In $\eta$ and $\zeta$, the factor of time arises in the factor $A_1 e^{\chi_1}+A_2 e^{\chi_2}$. Consider $A_1$, given by $\int e^{-\chi_1} dt$, and, $\chi_1 = -(t/\tau_R) -i(qB_0 /\epsilon_f) \int e^{-t/\tau_B} dt$. Taking the limit $qB_0 \rightarrow 0$, we see that,
\begin{equation}
   \int e^{(t/\tau_R) +i(qB_0 /\epsilon_f) \int e^{(-t/\tau_B)} dt} dt \rightarrow \int e^{(t/\tau_R) +i(0) \int e^{(-t/\tau_B)} dt} dt = \int e^{(t/\tau_R)} dt = \tau_R e^{t/\tau_R}
\end{equation}
Therefore, $A_1 e^{\chi_1}$ is, 
\begin{equation}
  \tau_R e^{t/\tau_R}e^{-t/\tau_R}=\tau_R.
\end{equation}
Similarly, $A_2 e^{\chi_2}=\tau_R$, and hence we have, $A_1 e^{\chi_1}+A_2 e^{\chi_2}=2\tau_R$. This is the same result one obtains in the $t\to \infty$ limit.

\section{Derivation of $\zeta$ [Eq.\eqref{zeta}]}\label{AppendixF1}
The initial expression of $\zeta$ is Eq.\eqref{41}:
\begin{equation}
  \zeta =\frac{1}{2} \sum_f \frac{g_f}{3} \int \frac{d^3 p}{(2 \pi)^3} \frac{p^2}{\epsilon_f}\left\{ f_0\left(1-f_0\right)A_f+\overline{f_0}\left(1-\overline{f_0}\right)\overline{A_f}\right\}+\frac{g_g}{3} \int \frac{d^3 p}{(2 \pi)^3} \frac{p^2}{\epsilon_g} f^g_0\left(1+f^g_0\right)A_g,\label{zeta1}  
\end{equation}
We look at the quark contribution to $\zeta$. The relevant landau Lifshitz conditions [Eqs.(\ref{43}-\ref{43a})] lead to
\begin{align}
 c_f&=\frac{\sum\limits_f g_f\int \frac{d^3 p}{(2 \pi)^3}\epsilon_f\,f_0\left(1-f_0\right)A_f}{T^2C_V}\label{cf},\\[0.3em]
 \overline{c_f}&=\frac{\sum\limits_f g_f\int \frac{d^3 p}{(2 \pi)^3}\epsilon_f\,\overline{f_0}\left(1-\overline{f_0}\right)\overline{A_f}}{T^2C_V}.
\end{align}
where we have used the definition of $C_V$ from Eq.\eqref{cv}. Thus, we have
\begin{align}
\zeta_q=\frac{1}{2} \sum_f \frac{g_f}{3} \int \frac{d^3 p}{(2 \pi)^3} \frac{p^2}{\epsilon_f}\left\{ f_0\left(1-f_0\right)A_f+\overline{f_0}\left(1-\overline{f_0}\right)\overline{A_f}\right\}-\frac{1}{2} \sum_fc_f \frac{g_f}{3} \int \frac{d^3 p}{(2 \pi)^3} p^2\left\{ f_0\left(1-f_0\right)+\overline{f_0}\left(1-\overline{f_0}\right)\right\}
\end{align}
Using Eq.\eqref{entropy}, this becomes
\begin{align}
\zeta_q=\frac{1}{2} \sum_f \frac{g_f}{3} \int \frac{d^3 p}{(2 \pi)^3} \frac{p^2}{\epsilon_f}\left\{ f_0\left(1-f_0\right)A_f+\overline{f_0}\left(1-\overline{f_0}\right)\overline{A_f}\right\}-\frac{1}{2} \frac{1}{3}(c_f+\overline{c_f})(3T^2s)
\end{align}
Substituting $c_f$ from Eq.\eqref{cf}, and using $v_s^2=s/c_v$, we get,
\begin{align}
 \zeta_q&=\frac{1}{2} \sum_f \frac{g_f}{3} \int \frac{d^3 p}{(2 \pi)^3} \frac{p^2}{\epsilon_f}\left\{ f_0\left(1-f_0\right)A_f+\overline{f_0}\left(1-\overline{f_0}\right)\overline{A_f}\right\}-\frac{1}{2}v_s^2 \sum_f g_f\int \frac{d^3 p}{(2 \pi)^3} \left\{\epsilon_f f_0\left(1-f_0\right)A_f+\overline{f_0}\left(1-\overline{f_0}\right)\overline{A_f}\right\}\nonumber\\
 &=\frac{1}{2}\sum_f \frac{g_f}{3}\int \frac{d^3 p}{(2 \pi)^3\epsilon_f}\left\{\epsilon_f f_0\left(1-f_0\right)A_f+\overline{f_0}\left(1-\overline{f_0}\right)\overline{A_f}\right\}\left[p^2-3v_s^2\epsilon_f^2\right].
\end{align}
Using $A_f = \frac{\beta}{3}(A_1 e^{\chi_1} +A_2 e^{\chi_2})\left[ \frac{p^2}{\epsilon_f}-3\epsilon_f\frac{\partial P}{\partial \epsilon}\right]$, $\overline{A_f} =\frac{\beta}{3} \left(\overline{A_1} e^{\overline{\chi_1}} +\overline{A_2} e^{\overline{\chi_2}}\right)\left[\frac{p^2}{\epsilon_f} -3\epsilon_f \frac{\partial P}{\partial \epsilon_f} \right]$, we finally arrive at
\begin{align}\label{zetaq}
    \zeta_q=\frac{\beta}{18}\sum_fg_f\int \frac{d^3 p}{(2 \pi)^3\epsilon_f^2}\left\{ f_0\left(1-f_0\right)\left(A_1e^{\chi_1}+A_2e^{\chi_2}\right)+\overline{f_0}\left(1-\overline{f_0}\right)\left(\overline{A_1}e^{\overline{\chi_1}}+\overline{A_2}e^{\overline{\chi_2}}\right)\right\}\left[p^2-3v_s^2\epsilon_f^2\right]^2.
\end{align}
The gluon contribution can be evaluated following similar steps. The result is
\begin{equation}\label{zetag}
  \zeta_g=  \frac{\beta g_g}{9} \int \frac{d^3 p}{(2 \pi)^3\,\epsilon_g^2}  \tau_g\,f^g_0\left(1+f^g_0\right)\left[p^2-3v_s^2\epsilon_g^2\right]^2.
\end{equation}
Adding Eq.\eqref{zetaq} and Eq.\eqref{zetag} gives us the total bulk viscosity of the medium, which is Eq.\eqref{zeta} in the main text.

\section{Derivation of Eqs. [\ref{etaa1}-\ref{etaa4}]}\label{AppendixD}
Comparing Eq.\eqref{dtij} and Eq.\eqref{dtbasis}, we get
\begin{align*}
    Y^1_{ij}\eta_1&= \frac{1}{15}\sum_f g_f \int \frac{d^3 p}{(2 \pi)^3} \epsilon_f \,v^4(\delta_{ij}\delta_{mn}+\delta_{im}\delta_{jn}+\delta_{in}\delta_{jn})C_1Y^1_{mn}\\
    &=\frac{1}{15}\sum_f g_f \int \frac{d^3 p}{(2 \pi)^3} \epsilon_f \,v^4C_1(Y^1_{ii}+Y^1_{ij}+Y^1_{ji}).
\end{align*}
$Y_{ii}=0$, and $Y_{ij}=Y_{ji}$ yields
\begin{equation}
    \eta_1 = \frac{2}{15} \sum_f g_f \int \frac{d^3 p}{(2 \pi)^3} \epsilon_f\, v^4 C_1.
\end{equation}
Similarly, $\eta_2$, $\eta_3$, and $\eta_4$ are derived.

\section{Derivation of $\eta_i$ [Eqs. \ref{eta1}-\ref{eta4}]}\label{AppendixE}
In the co-moving frame, $\partial^{i}f_0|_{\bm{V}=0}=-\beta f_0(1-f_0)p_{j}\partial^{i}V^{j}$, where $f_0$ is the F-D distribution function. Using this, the Boltzmann equation becomes
\begin{equation}\label{begen}
    -v_iv_j\epsilon_f\beta f_0(1-f_0)V_{ij}+qE^i\frac{\partial f^0}{\partial p^i}+\omega_cv_jb_{ij}\frac{\partial}{\partial v^i}(\delta f)+\frac{\partial}{\partial t}(\delta f)=-\frac{\delta f}{\tau_R}.
\end{equation}
With $\delta f$ being given by Eq.\eqref{50}, the above equation reads 
\begin{equation}
    - \beta \epsilon_f V_{ij} v_i v_j f_0 (1-f_0)+qE^i\frac{\partial f^0}{\partial p^i}+\omega_c b_{ij}v_j \frac{\partial}{\partial v_i} \bigg(\sum_{l=0}^4 C_l Y_{mn}^l v_m v_n \bigg)+\sum_{l=0}^4 \dot{C}_l Y_{mn}^l v_m v_n = -\frac{\sum_{l=0}^4 C_l Y_{mn}^l v_m v_n}{\tau_R}.  
\end{equation}
This is Eq.\eqref{begeneral}. With the relations $V_{ij} b_i b_j = 0$, $b_{ij} v_i v_j = 0$, $b_i b_i = 1$, $b_{ij} b_i = 0$, and $b_{ij} b_j = 0$, the basis tensors $Y^l_{mn}$ reduce to
\begin{align}
    Y^1_{mn}&=2V_{mn}-2V_{mk}\, b_kb_n-2V_{nk}\,b_kb_m\\
    Y^2_{mn}&=2V_{mk}\, b_kb_n+2V_{nk}\,b_kb_m\\
    Y^3_{mn}&=V_{mk}\,b_{nk}+V_{nk}\,b_{mk}-V_{kl}\,b_{mk}b_nb_l-V_{kl}\,b_{nk}b_mb_l\\
    Y^4_{mn}&=2V_{kl}\,b_{mk}b_nb_l+2V_{kl}\,b_{nk}b_mb_l.
\end{align}
We now simplify the terms in the L.H.S. of Eq.\eqref{begen} containing derivatives of $\delta f$. Using the above expressions, we can write
\begin{align}
    v_jb_{ij}\frac{\partial}{\partial v^i}(\delta f)=&2v_jb_{ij}\Big[C_1(2V_{mi}\,v_m-\underline{2V_{mk}\,b_kb_iv_m}-2V_{ik}\,b_kb_mv_m)+C_2(\underline{2V_{mk}\,b_kb_iv_m}+2V_{ik}\,b_kb_mv_m)+\nonumber\\&C_3(V_{mk}\,b_{ik}\,v_m + V_{ik}b_{mk}v_m - \underline{V_{kl}b_{mk}b_ib_lv_m}-V_{kl}b_{ik}b_mb_lv_m)+C_4(\underline{2V_{kl}b_{mk}b_ib_lv_m} + 2V_{kl}b_{ik}b_mb_lv_m)\Big]
\end{align}
The underlined terms vanish owing to $b_{ij}b_i=0$. Thus, 
\begin{align}
    v_jb_{ij}\frac{\partial}{\partial v^i}(\delta f)=2\omega_c\Big[2C_1V_{ik}b_{ij}v_jv_k&-2C_1V_{ik}b_{ij}b_kv_j(\bm{b}\cdot \bm{V})+2C_2V_{ik}b_{ij}b_kv_j(\bm{b}\cdot \bm{V})\nonumber\\&+2C_3V_{ij}v_iv_j-4C_3V_{ij}b_iv_j(\bm{b}\cdot \bm{V})+2C_4V_{ij}b_iv_j(\bm{b}\cdot \bm{V})\Big]
\end{align}
Next, we simplify $\delta f$ 
\begin{align}
    \delta f=&2C_1v_mv_n(V_{mn}-V_{mk}b_kb_n-V_{nk}b_kb_m)+2C_2v_mv_n(V_{mk}b_kb_n+V_{nk}b_kb_m)+\nonumber\\&C_3v_mv_n(V_{mk}b_{nk}+V_{nk}b_{mk}-V_{kl}b_{mk}b_nb_l-V_{kl}b_{nk}b_mb_l)+2C_4v_mv_n(V_{kl}b_{mk}b_nb_l+V_{kl}b_{nk}b_mb_l)\nonumber\\[0.5em]
    =&2C_1v_iv_jV_{ij}-4C_1v_ib_jV_{ij}(\bm{v}\cdot \bm{b})+4C_2v_ib_jV_{ij}(\bm{v}\cdot \bm{b})+C_3V_{ik}b_{jk}v_iv_j+\nonumber\\&C_3V_{jk}b_{ik}v_iv_j-2C_3V_{kl}b_{ik}b_lv_i(\bm{v}\cdot \bm{b})+4C_4V_{kl}b_{ik}b_lv_i(\bm{v}\cdot \bm{b})
\end{align}
Similarly we write out the term $\frac{\partial}{\partial t}(\delta f)$ explicitly. Then, the Boltzmann equation becomes
\begin{multline}
  \beta \epsilon_f V_{ij} v_i v_j f_0 (1-f_0)+qE^i\frac{\partial f^0}{\partial p^i}+2\omega_c\Big[2C_1V_{ik}b_{ij}v_jv_k-2C_1V_{ik}b_{ij}b_kv_j(\bm{v}\cdot \bm{b})+2C_2V_{ik}b_{ij}b_kv_j(\bm{v}\cdot \bm{b})+2C_3V_{ij}v_iv_j-\\
  4C_3V_{ij}b_iv_j(\bm{v}\cdot \bm{b})+2C_4V_{ij}b_iv_j(\bm{v}\cdot \bm{b})\Big]+2C_1v_iv_jV_{ij}-4\dot{C_1}(\bm{v}\cdot \bm{b})V_{ij}v_ib_j +4\dot{C_2}(\bm{v}\cdot \bm{b})V_{ij}v_ib_j+2\dot{C_3}V_{ik}b_{jk}v_iv_j -\\
  2\dot{C_3}(\bm{v}\cdot \bm{b})V_{kl}b_{ik}b_lv_i+4\dot{C_4}(\bm{v}\cdot \bm{b})V_{kl}b_{ik}b_lv_i=\frac{-1}{\tau_R}\Big[2C_1v_iv_jV_{ij}-4C_1v_ib_jV_{ij}(\bm{v}\cdot \bm{b})+4C_2v_ib_jV_{ij}(\bm{v}\cdot \bm{b})+\\
  C_3V_{ik}b_{jk}v_iv_j+C_3V_{jk}b_{ik}v_iv_j-2C_3V_{kl}b_{ik}b_lv_i(\bm{v}\cdot \bm{b})+4C_4V_{kl}b_{ik}b_lv_i(\bm{v}\cdot \bm{b})\Big]. 
\end{multline}
Comparing tensor structures on both sides of the above equation, we get,
\begin{align}
  -\epsilon_f\beta f_0(1-f_0)+4C_3\omega_c+2\dot{C_1}&=-\frac{1}{\tau_R}2C_1\\
  \left(-8C_3\omega_c+4\dot{C_2}+4\omega_cC_4-4\dot{C_1}\right)&=-\frac{1}{\tau-R}\left(4C_2-4C_1\right)\\
  \dot{C_3}-2\omega_cC_1&=-\frac{1}{\tau_R}C_3\\
  -4\omega_cC_1+4\omega_cC_2+2\dot{C_3}-4\dot{C_4}&=-\frac{1}{\tau-R}\left(2C_3-4C_4\right)
\end{align}
This leads to the following set of coupled differential equations
\begin{align}
    \dot{C_1}&=-\frac{C_1}{\tau_R}-2\omega_cC_3+\frac{\epsilon_f\beta}{2}f_0(1-f_0)\\
    \dot{C_3}&=-\frac{C_3}{\tau_R}-2\omega_cC_1\\
    \dot{C_4}&=-\frac{C_4}{\tau_R}+\omega_cC_2\\
    \dot{C_2}&=-\frac{C_2}{\tau_R}-\omega_cC_4+\frac{\epsilon_f\beta}{2}f_0(1-f_0)
\end{align}
The first two equations can be written as a matrix differential equation
\begin{equation}
    \frac{d\bm{X_1}}{dt}=G_1(t)\bm{X_1}+\bm{H_1}(t),\quad \text{with}
    \end{equation}
\begin{equation}
		\bm{X_1}\equiv\begin{pmatrix}
			C_1\\
			C_3
		\end{pmatrix},\quad G_1(t)\equiv \begin{pmatrix}
		-1/\tau_R & -2\omega_c\\
		2\omega_c & -1/\tau_R
	\end{pmatrix},\quad \bm{H_1}(t)=\begin{pmatrix}
	\frac{\epsilon_f\beta}{2}f_0(1-f_0)\\
	0
\end{pmatrix}
	\end{equation}
Similarly, the last two equations can also be written as a matrix differential equation  
\begin{equation}
    \frac{d\bm{X_2}}{dt}=G_2(t)\bm{X_2}+\bm{H_2}(t),\quad \text{with}
    \end{equation}
\begin{equation}
		\bm{X_2}\equiv\begin{pmatrix}
			C_2\\
			C_4
		\end{pmatrix},\quad G_2(t)\equiv \begin{pmatrix}
		-1/\tau_R & -\omega_c\\
		\omega_c & -1/\tau_R
	\end{pmatrix},\quad \bm{H_2}(t)=\begin{pmatrix}
	\frac{\epsilon_f\beta}{2}f_0(1-f_0)\\
	0
\end{pmatrix}
	\end{equation}
The solutions to these differential equations have a complementary function [Eq.\eqref{cf}] and a particular integral [Appendix \eqref{AppendixB}]. Following the same steps as enumerated earlier, we finally arrive at
\begin{align}
    &C_1 = \frac{\epsilon_f \beta f^0 (1-f^0)}{4}\bigg[\bigg(\int e^{-\xi_1}dt  \bigg)e^{\xi_1}+\bigg(\int e^{-\xi_2}dt  \bigg)e^{\xi_2}\bigg],\\
    &C_2 = \frac{\epsilon_f \beta f^0 (1-f^0)}{4}\bigg[\bigg(\int e^{-\xi_3}dt  \bigg)e^{\xi_3}+\bigg(\int e^{-\xi_4}dt\bigg)e^{\xi_4}\bigg],\\
     &C_3 = \frac{i\epsilon_f \beta f^0 (1-f^0)}{4}\bigg[\bigg(\int e^{-\xi_1}dt  \bigg)e^{\xi_1}-\bigg(\int e^{-\xi_2}dt\bigg)e^{\xi_2}\bigg],\\
     &C_4 = \frac{i\epsilon_f \beta f^0 (1-f^0)}{4}\bigg[\bigg(\int e^{-\xi_3}dt  \bigg)e^{\xi_3}-\bigg(\int e^{-\xi_4}dt\bigg)e^{\xi_4}\bigg],
\end{align}
with
\begin{align}
    &\xi_1 = -\frac{t}{\tau_R} -2i\int\frac{qB}{\epsilon_f}dt,\\
    &\xi_2 = -\frac{t}{\tau_R} +2i\int\frac{qB}{\epsilon_f}dt,\\
    &\xi_3 = -\frac{t}{\tau_R} -i\int\frac{qB}{\epsilon_f}dt,\\
    &\xi_4 = -\frac{t}{\tau_R} +i\int\frac{qB}{\epsilon_f}dt
\end{align}
These are Eqs.[\ref{c1}-\ref{xi4}] in the main text. Now substituting the $C_i$'s thus obtained in Eqs.[\ref{etaa1}-\ref{etaa4}], we arrive at
\begin{align}
    &\eta_1 = \frac{1}{60 \pi^2} \sum_f g_f \int dp \frac{p^6}{\epsilon_f^2}\beta f^0 (1-f^0)\bigg[e^{\xi_1}\int e^{-\xi_1}dt+e^{\xi_2} \int e^{-\xi_2}dt\bigg],\\
    &\eta_2 = \frac{1}{60 \pi^2} \sum_f g_f \int dp \frac{p^6}{\epsilon_f^2}\beta f^0 (1-f^0)\bigg[e^{\xi_3}\int e^{-\xi_3}dt+e^{\xi_4} \int e^{-\xi_4}dt\bigg],\\
    &\eta_3 = \frac{i}{60 \pi^2} \sum_f g_f \int dp \frac{p^6}{\epsilon_f^2}\beta f^0 (1-f^0)\bigg[e^{\xi_1}\int e^{-\xi_1}dt-e^{\xi_2} \int e^{-\xi_2}dt\bigg],\\
    &\eta_4 = \frac{i}{60 \pi^2} \sum_f g_f \int dp \frac{p^6}{\epsilon_f^2}\beta f^0 (1-f^0)\bigg[e^{\xi_3}\int e^{-\xi_3}dt-e^{\xi_4} \int e^{-\xi_4}dt\bigg].
\end{align}
These are Eqs.[\ref{eta1}-\ref{eta4}] in the main text.
 
\end{appendices}

\bibliography{ref}{}

\end{document}